\documentclass[12pt]{article}
\usepackage{amssymb,amsmath,amsbsy}
\usepackage{dsfont}
\usepackage[title]{appendix}
\usepackage{hyperref}
\makeatletter
\g@addto@macro\bfseries{\boldmath}
\makeatother
\DeclareMathOperator{\Tr}{Tr}
\newenvironment{Eqnarray}%
     {\arraycolsep 0.14em\begin{eqnarray}}{\end{eqnarray}}

\renewcommand{\Re}{\operatorname{Re}}
\renewcommand{\Im}{\operatorname{Im}}
\newcommand{\sgn}{\operatorname{sgn}}

\def\half{\tfrac{1}{2}}
\def\beqa{\begin{Eqnarray}}
\def\eeqa{\end{Eqnarray}}
\def\beq{\begin{equation}}
\def\eeq{\end{equation}}
\def\eq#1{eq.~(\ref{#1})}
\def\eqst#1#2{eqs.~(\ref{#1})--(\ref{#2})}
\def\eqss#1#2#3{eqs.~(\ref{#1}), (\ref{#2}) and (\ref{#3})}
\def\eqs#1#2{eqs.~(\ref{#1}) and (\ref{#2})}
\def\Eq#1{Eq.~(\ref{#1})}
\def\eqs#1#2{eqs.~(\ref{#1}) and (\ref{#2})}
\def\Eqs#1#2{Eqs.~(\ref{#1}) and (\ref{#2})}
\def\Eqss#1#2#3{Eqs.~(\ref{#1}), (\ref{#2}) and (\ref{#3})}
\def\llsup#1{^{\lower 2pt\hbox{$\scriptstyle#1$}}}
\def\ifmath#1{\relax\ifmmode #1\else $#1$\fi}
\def\half{\ifmath{{\tfrac12}}}

\def\ls#1{\ifmath{_{\lower1.5pt\hbox{$\scriptstyle #1$}}}}
\def\newcdot{\kern.06em{\cdot}\kern.06em}
\def\phm{\phantom{-}}

\def\pmat#1{\begin{pmatrix}#1\end{pmatrix}}

\def\T{{\mathsf T}}

\def\ip#1#2{\langle #1|#2\rangle}

\def\id{{\mathds{1}_{2\times 2}}}

\def\lsup#1{^{\lower 3pt\hbox{$\scriptstyle#1$}}}
\catcode`~=11 
\newcommand{\urltilde}{\kern -.15em\lower .7ex\hbox{~}\kern .04em}
\catcode`~=13 

\begin{document}
\begin{center}
{\Large \bf  A tale of three diagonalizations}\\[1cm]
{\Large Howard E. Haber}\\[5pt]
{\large Santa Cruz Institute for Particle Physics  \\[4pt]
University of California, Santa Cruz, CA 95064, USA } 
\end{center}
\vskip 0.2in

\begin{abstract}
In addition to the diagonalization of a normal matrix by a unitary similarity transformation, there are two other types
of diagonalization procedures that sometimes arise in quantum theory applications---the singular value decomposition
and the Autonne-Takagi factorization.   In this pedagogical review, each of these diagonalization procedures is performed
for the most general $2\times 2$ matrices for which the corresponding diagonalization is possible, and explicit
analytical results are provided in each of the three cases.
\end{abstract}

\section{Introduction}

In quantum physics, some problems can be reduced to two state systems.   The solution to these problems involves
the diagonalization of the $2\times 2$ hermitian matrix Hamiltonian $H$, which consists of reducing $H$ via a unitary similarity transformation to a diagonal
matrix whose elements are the (real) eigenvalues of $H$. Instead of repeating the diagonalization every  time
a problem of this type arises, it is convenient to solve it once and for all by considering the diagonalization of a general
$2\times 2$ hermitian matrix.  In fact, it is possible to be slightly more general.   Recall that a matrix is normal (i.e.~the matrix
commutes with its hermitian adjoint) if and only if it is diagonalizable by a unitary similarity transformation (see, e.g., Theorem 2.5.3 of Ref.~\cite{horn}).  Hence, 
this pedagogical review will begin by providing the explicit diagonalization of a general $2\times 2$ normal matrix.

Two additional diagonalization procedures often arise in the quantum field theories of fermions (see, e.g., Ref.~\cite{Dreiner:2008tw}).  The fermion mass eigenstates
are identified by reducing the fermion mass matrix to diagonal form.  But, in such problems, the relevant diagonalization procedure
is not carried out by a unitary similarity transformation.   In general, the mass matrix that arises in a theory of charged fermions is a complex
matrix with no other special features.  The relevant diagonalization procedure is called the singular value decomposition of a complex matrix (see, e.g., Refs.~\cite{horn,horn2}).
This decomposition produces a diagonal matrix whose diagonal elements are real and nonnegative, corresponding
to the physical masses of the charged fermions.   In contrast, the mass matrix that arises in a theory of neutral (Majorana) fermions is a complex
symmetric matrix.  The relevant diagonalization procedure is called the Autonne-Takagi factorization of a complex symmetric matrix~\cite{autonne,takagi}.
This factorization also produces a diagonal matrix whose diagonal elements are real and nonnegative, corresponding
to the physical masses of the neutral fermions. 

In this review, we apply the three diagonalization procedures mentioned above to a complex normal matrix, an arbitrary complex matrix, and a complex symmetric matrix, 
respectively.  In each case, we diagonalize the corresponding $2\times 2$ matrix explicitly and provide 
analytic results for the corresponding
diagonalizing matrix and for the elements of the resulting diagonal matrix.

\section{The diagonalization of a \texorpdfstring{$2\times 2$}{2x2} normal matrix by a unitary similarity transformation}

 \label{sec:normal}
 \bigskip
 
 Consider a general $2\times 2$ complex matrix,
 \beq \label{Nmatrix}
 N=\left(\begin{array}{cc} a\, &\quad b\\ c\, &\quad d\end{array}\right)\,.
\eeq
Then, $N$ is normal if 
\beq \label{Nnormal}
N^\dagger N=NN^\dagger\,.
\eeq
Inserting \eq{Nmatrix} into \eq{Nnormal}, it follows that\footnote{\Eqs{Imdma}{ABrelation} have been inspired by Problem 2.5.P29 of Ref.~\cite{horn}.} 
\beq  \label{Imdma}
|b|=|c|\,,\qquad \quad \Im\bigl[(d-a)e^{-i(\alpha+\beta)/2}\bigr]=0\,,
 \eeq
 where
 \beq
 \alpha\equiv\arg b\,,\qquad\quad \beta\equiv\arg c\,.
 \eeq
 It is then straightforward to verify that the matrix 
\beq \label{ABrelation}
A=e^{-i(\alpha+\beta)/2}(N-a\mathds{1}_{2\times 2})=\begin{pmatrix} 0 & \quad |b|e^{i(\alpha-\beta)/2} \\  |b|e^{-i(\alpha-\beta)/2} & \quad (d-a)e^{-i(\alpha+\beta)/2}\end{pmatrix} \,,
\eeq
is hermitian, where $\mathds{1}_{2\times 2}$ is the $2\times 2$ identity matrix.

The diagonalization of $N$ by a unitary similarity transformation is given by,
\beq \
U^{-1}NU=\begin{pmatrix} \mu_1 & \quad 0 \\ 0 & \quad \mu_2\end{pmatrix}\,,
\eeq
where $\mu_1$ and $\mu_2$ are the complex eigenvalues of $N$,
\beq \label{mueigen}
\mu_{1,2}=\frac12\left[a+d\mp\sqrt{(a-d)^2+4|b|^2e^{i(\alpha+\beta)}}\right]\,.
\eeq
Using \eq{ABrelation}, it follows that
\beq \label{UNU}
U^{-1}NU=e^{i(\alpha+\beta)/2} U^{-1}A U+a\mathds{1}_{2\times 2}\,.
\eeq
Hence, to diagonalize $N$, we must diagonalize the hermitian matrix $A$.   We will carry out this procedure in Section~\ref{diaghermitian}, which will provide an explicit expression for the diagonalizing matrix~$U$.

The eigenvalues of an hermitian matrix are real.  Denoting the eigenvalues of $A$ by $\lambda_1$ and $\lambda_2$, one easily obtains
\beq
\lambda_{1,2} = \frac{1}{2}\left[(d-a)e^{-i(\alpha+\beta)/2} \mp \sqrt{\bigl[(d-a)e^{-i(\alpha+\beta)/2}\bigr]^2 + 4|b|^2}\right]\,. 
\eeq
Note that in light of \eq{Imdma}, it follows that $\lambda_1$ and $\lambda_2$ are real numbers.
Hence, \eq{UNU} yields,
\beq \label{mu12}
\mu_{1,2}=e^{i(\alpha+\beta)/2}\lambda_{1,2}+a\,.
\eeq
It is straightforward to 
check that \eqs{mu12}{mueigen} are equivalent.

 \section{The diagonalization of a \texorpdfstring{$2\times 2$}{2x2} hermitian matrix by a unitary similarity transformation}
 \label{diaghermitian}
 \bigskip

Consider a general $2\times 2$ hermitian matrix
\beq \label{Amatrix}
A=\left(\begin{array}{cc} a\, &\,\,\, c\\ c^\ast \, &\,\,\, b\end{array}\right)\,,
\eeq
where $a$ and $b$ are real numbers and the complex number $c$ expressed in polar exponential form is given by,
\beq \label{cdef}
c=|c|e^{i\phi}\,,\qquad \text{where $0\leq\phi<2\pi$}\,.
\eeq

The eigenvalues are the roots of the characteristic equation:
\beq
\det \begin{pmatrix} a -\lambda &\,\,\,c \\c^\ast&\,\,\,b-\lambda\end{pmatrix} = (a-\lambda)(b-\lambda) - |c|^2 = \lambda^2 - \lambda(a+b) + (ab-|c|^2)
= 0 \,.
\eeq
Noting that $(a+b)^2-4(ab-|c|^2)=(a-b)^2+4|c|^2$, the two roots can be written as:
\beq \label{roots}
\lambda_1 = \tfrac{1}{2}\left[a+b - \sqrt{(a-b)^2 + 4|c|^2}\right] \quad \text{and} \quad
     \lambda_2 = \tfrac{1}{2}\left[a+b + \sqrt{(a-b)^2 + 4|c|^2}\right]\,,
\eeq
where by convention we take $\lambda_1 \leq \lambda_2$.  As expected, the eigenvalues of the hermitian matrix $A$ are real.

An hermitian matrix can be diagonalized by a unitary matrix $U$,
\beq \label{UAU}
U^{-1}AU=\left(\begin{array}{cc} \lambda_1\, & 0\\ 0 \,& \lambda_2\end{array}\right)\,,
\eeq
where $\lambda_1$ and $\lambda_2$ are the eigenvalues obtained in \eq{roots}.
Note that one can always transform $U\to e^{i\zeta}U$ without modifying \eq{UAU}, since the phase factor cancels out.
Since $\det U$ is a complex number of unit modulus, one can choose $\det U=1$ in \eq{UAU} without loss of generality.
The most general $2\times 2$ unitary matrix of unit determinant can be written as,
$$
U=\left(\begin{array}{cc} \phm e^{i\beta}\,\cos\theta\,&\,\,\, e^{i\chi}\sin\theta\\ -e^{-i\chi}\sin\theta 
\, &\,\,\, e^{-i\beta}\cos\theta\ \end{array}\right)\,.
$$
The columns of $U$ are the normalized eigenvectors of $A$ corresponding to the eigenvalues $\lambda_1$
and $\lambda_2$, respectively.  But, we are always free to multiply any normalized eigenvector
by an arbitrary complex phase factor.  Thus, without loss of generality, we can choose $\beta=0$ and $\cos\theta\geq 0$.
Moreover, the sign of $\sin\theta$ can always be absorbed into the definition of $\chi$.  Hence,
we will take
\beq \label{Udef}
U=\left(\begin{array}{cc} \cos\theta\,&\,\,\, e^{i\chi}\sin\theta\\ -e^{-i\chi}\sin\theta 
\, &\,\,\, \cos\theta\ \end{array}\right)\,,
\eeq
where
\beq \label{range}
0\leq\theta\leq\half\pi\,,\quad\text{and}\quad 0\leq\chi<2\pi\,.
\eeq

We now plug in \eq{Udef} into \eq{UAU}.
Since the off-diagonal terms must vanish, one obtains constraints on the angles $\theta$ and $\chi$.
In particular,
\beqa
U^{-1} A U
&=& \pmat{\cos\theta\quad & -e^{i\chi}\sin\theta\\e^{-i\chi}\sin\theta\quad& \cos\theta}\pmat{a\quad&|c|e^{i\phi}\\|c|e^{-i\phi}\quad&b}
        \pmat{\cos\theta\quad&e^{i\chi}\sin\theta\\ -e^{-i\chi}\sin\theta\quad& \cos\theta} \nonumber \\[8pt] 
&=&\pmat{\cos\theta\quad & -e^{i\chi}\sin\theta\\e^{-i\chi}\sin\theta\quad& \cos\theta}
    \pmat{a\cos\theta - |c|e^{i(\phi-\chi)}\sin\theta\quad&ae^{i\chi}\sin\theta+|c|e^{i\phi}\cos\theta\\ |c|e^{-i\phi}\cos\theta-be^{-i\chi}\sin\theta&\quad
|c|e^{-i(\phi-\chi)}\sin\theta+b\cos\theta} \nonumber\\[8pt]
&=& \pmat{\lambda_1\quad&Z\\Z^\ast\quad&\lambda_2} \,,\nonumber
\eeqa
where
\beqa
\lambda_1&=&a\cos^2\theta-2|c|\cos\theta\sin\theta\cos(\phi-\chi)+b\sin^2\theta\,,\label{lam1} \\[6pt]
\lambda_2&=&a\sin^2\theta+2|c|\cos\theta\sin\theta\cos(\phi-\chi)+b\cos^2\theta\,,\label{lam2} \\[6pt]
Z&=&
e^{i\chi}\biggl\{(a-b)\cos\theta\sin\theta +|c|\Bigl[e^{i(\phi-\chi)}\cos^2\theta-e^{-i(\phi-\chi)}\sin^2\theta\Bigr]\biggr\}\,.\label{Z}
\eeqa
The vanishing of the off-diagonal elements of $U^{-1} A U$ implies that:
$$
(a-b)\cos\theta\sin\theta +|c|\Bigl[e^{i(\phi-\chi)}\cos^2\theta-e^{-i(\phi-\chi)}\sin^2\theta\Bigr]=0\,.
$$
This is a complex equation.  Taking real and imaginary parts yields two real equations,
\beqa
&&\half(a-b)\sin 2\theta+|c|\cos 2\theta\cos(\phi-\chi)=0\,,\label{Rpart}\\[6pt]
&&|c|\sin(\phi-\chi)=0\,.\label{Ipart}
\eeqa

Consider first the special case of $c=0$.  Then, in light of our convention that $\lambda_1\leq\lambda_2$,
\beqa
&&c=0 \quad {\rm and} \quad a<b\quad\Longrightarrow\quad \theta=0\,\,\,\,\,\,\,\text{and $\chi$ is undefined}\,,\nonumber\\[6pt] 
&&c=0 \quad {\rm and} \quad a>b\quad\Longrightarrow\quad \theta=\half\pi\,\,\, \text{and $\chi$ is undefined}\,,\nonumber\\[6pt] 
&&c=0 \quad {\rm and} \quad a=b\quad\Longrightarrow\quad \text{$\theta$ and $\chi$ are undefined}\,.\nonumber
\eeqa
In particular, if $c=0$ and $a=b$, then $A=a\id$ and it follows that $U^{-1}AU=U^{-1}U=a\id$, which is satisfied for any unitary matrix $U$.
Consequently, in this limit $\theta$ and $\chi$ are arbitrary and hence undefined, as indicated above.    

If $c\neq 0$ then \eq{Ipart} yields 
\beq \label{twosigns}
\sin(\phi-\chi)=0\,\quad \text{and}\quad \cos(\phi-\chi)=\varepsilon\,,\quad \text{where $\varepsilon =\pm 1$}.
\eeq
We can determine the sign $\varepsilon$ as follows.   Since $\lambda_1\leq\lambda_2$, we subtract \eqs{lam1}{lam2} and make use of \eq{twosigns} to obtain,
\beq \label{inequality}
(a-b)\cos 2\theta -2\varepsilon |c|\sin 2\theta\geq 0\,.
\eeq
Likewise, we insert \eq{twosigns} into \eq{Rpart}, which yields
\beq \label{equality}
(a-b)\sin 2\theta+2\varepsilon|c|\cos 2\theta=0\,.
\eeq
Finally, we multiply \eq{inequality} by $\sin 2\theta$ and \eq{equality} by $\cos 2\theta$ and subtract the two resulting equations.  The end result is,
\beq
2\varepsilon |c|\geq 0\,.
\eeq
By assumption, $c\neq 0$.  Thus, it follows that $\varepsilon\geq 0$.  Since $\varepsilon=\pm 1$, we can conclude that $\varepsilon =1$.
Hence,
\beq \label{varepsone}
\cos(\phi-\chi)=1\,,\quad \text{for $c\neq 0$}.
\eeq
By the conventions established in \eqs{cdef}{range}, we take $0\leq\phi\,,\,\chi< 2\pi$.  Hence, it follows that 
\beq \label{chiresult}
\chi=\phi\,.
\eeq

We can now determine $\theta$.  Inserting \eq{varepsone} into \eq{Rpart} yields
\beqa \label{angle}
\tan 2\theta = \frac{2|c|}{b-a}\,,\quad \text{for $c\neq 0$ and $a\neq b$}\,.
\eeqa
Note that if $a=b$, then \eq{equality} yields $\cos 2\theta=0$.  In light of \eq{range},
\beq \label{border}
c\neq 0 \quad {\rm and} \quad a=b \quad \Longrightarrow\quad \theta=\tfrac{1}{4}\pi\,.
\eeq
If $c\neq 0$ and $a\neq b$, then we can use \eq{angle} with the convention that $\sin 2\theta\geq 0$ [cf.~\eq{range}] to conclude that
\beqa 
\sin 2\theta&=&\frac{2|c|}{\sqrt{(b-a)^2+4|c|^2}}\,.\label{s2t} \\[8pt]
\cos 2\theta&=&\displaystyle\frac{b-a}{\sqrt{(b-a)^2+4|c|^2}}\,.\label{quad2}
\eeqa
Using the well known identity, $\tan\theta=(1-\cos 2\theta)/\sin 2\theta$, it follows that 
\beq \label{ttheta}
\tan\theta=\frac{a-b+\sqrt{(b-a)^2+4|c|^2}}{2|c|}\,,
\eeq
which is manifestly positive.  
It then follows that,
\beq \label{sctheta}
\sin\theta=\left(\frac{a-b+\sqrt{(b-a)^2+4|c|^2}}{2\sqrt{(b-a)^2+4|c|^2}}\right)^{1/2}\!\!,\qquad\quad
\cos\theta=\left(\frac{b-a+\sqrt{(b-a)^2+4|c|^2}}{2\sqrt{(b-a)^2+4|c|^2}}\right)^{1/2}\!\!.
\eeq
Indeed, the above results imply that the sign of $b-a$ determines whether $0<\theta<\tfrac{1}{4}\pi$
or $\tfrac{1}{4}\pi<\theta<\half\pi$.  The former corresponds to $a<b$ while the latter
corresponds to $a>b$.  The borderline case of $a=b$ has already been treated in \eq{border}.

To summarize, if $c\neq 0$, then eqs.~(\ref{chiresult}), (\ref{s2t}) and (\ref{quad2}) uniquely specify the diagonalizing matrix $U$
[in the conventions specified in \eqs{cdef}{range}].
When $c=0$ and $a\neq b$, it follows that $\chi$ is arbitrary and $\theta=0$ or $\half\pi$
for the two cases of $a<b$ or $a>b$, respectively.\footnote{Note that in the case of $c=0$ and $a>b$, the matrix $A$ is diagonal.   Nevertheless,
the ``diagonalizing'' matrix, $U\neq \id$.  Indeed, in this case $\theta=\half\pi$, and $U^{-1}AU$ simply interchanges
the two diagonal elements of~$A$ to ensure that $\lambda_1\leq \lambda_2$ in \eq{UAU}, as required by the convention adopted below \eq{roots}.}
Finally, if $c=0$ and $a=b$, then $A=a\id$,
in which case $U$ is arbitrary.
\clearpage

 \section{The diagonalization of a \texorpdfstring{$2\times 2$}{2x2} real symmetric matrix by an orthogonal similarity transformation}
 \label{diagortho}
 \bigskip

In this section, we consider a special case of the one treated in Section~\ref{diaghermitian} in which the matrix $A$ given in \eq{Amatrix} is real.  
That is, $c=c^*$, in which case $A$ is a real symmetric matrix that can be diagonalized 
by a real orthogonal matrix.  The two eigenvalues are still given by \eq{roots}
in the convention that $\lambda_1\leq\lambda_2$,
although the absolute values signs are no longer needed since for real values of $c$, we have $|c|^2=c^2$.
Moreover, since $c$ is real, \eq{cdef} implies that
if $c\neq 0$ then $\phi=0$ or $\phi=\pi$.  \Eq{chiresult} then yields 
\beq \label{realchi}
\chi=\begin{cases} 0\,,&\quad \text{for $c\neq 0$ and $\phi=0$}\,,\\ \pi\,,&\quad \text{for $c\neq 0$ and $\phi=\pi$}\,,\end{cases}
\eeq
which is equivalent to the statement that
\beq
e^{i\chi}={\rm sgn}~\!c\,,\quad \text{for real $c\neq 0$}.
\eeq
It is convenient to redefine $\theta\to \theta\,\sgn c$ in \eq{Udef}.   With this modification, the range of $\theta$ can be taken as\footnote{Using
$\cos(\theta+\pi)=-\cos\theta$ and $\sin(\theta+\pi)=-\sin\theta$, 
it follows that shifting $\theta\to\theta+\pi$ simply multiplies $U$ by an overall factor of $-1$.
In particular, $U^{-1}AU$ is unchanged.  Hence, the convention $-\half\pi<\theta\leq\half\pi$ may be chosen without loss of generality.}   
\beq \label{range2}
-\half\pi <\theta \leq \half\pi\,.
\eeq
The diagonalizing matrix $U$ is now
a real orthogonal $2\times 2$ matrix,
\beq \label{Udiag}
U=\left(\begin{array}{cc} \phm\!\! \cos\theta\,&\,\,\,\!\sin\theta\\ \!\!-\sin\theta 
 &\,\,\,\cos\theta\ \end{array}\right)\,, \qquad \text{where\quad $\Biggl\{\begin{matrix} \phm  c>0\quad &\Longrightarrow & \!\!\!\!\!\! \phantom{-\half}\,\, 0<\theta<\half\pi\,, \\
\phm  c=0\quad & \Longrightarrow &  \!\!\!\!\!\! \text{\phantom{xxxx}$\theta=0$ or $\theta=\half\pi$}\,, \\
\phm c<0\quad &\Longrightarrow &  \!\!\!\!\!\!  -\half\pi<\theta<0\,.\end{matrix}$}
\eeq

Hence, for real $c\neq 0$ with the range of $\theta$ specified in \eq{range2}, we see that
eqs.~(\ref{angle}) and (\ref{s2t})--(\ref{ttheta})
are modified by replacing $|c|$ with $c$.
For example,
\beq
\sin 2\theta=\frac{2c}{\sqrt{(b-a)^2+4c^2}}\,,\qquad\quad
\cos 2\theta=\displaystyle\frac{b-a}{\sqrt{(b-a)^2+4c^2}}\,.\label{quad4}
\eeq
It then follows that 
\beq \label{sctheta2}
\sin\theta=\sgn(c)\left(\frac{a-b+\sqrt{(b-a)^2+4c^2}}{2\sqrt{(b-a)^2+4c^2}}\right)^{1/2}\!\!,\qquad
\cos\theta=\left(\frac{b-a+\sqrt{(b-a)^2+4c^2}}{2\sqrt{(b-a)^2+4c^2}}\right)^{1/2}\!\!.
\eeq

The sign of $c$ determines the quadrant in which $\theta$ lives.  Moreover,
for $c> 0$, the sign of $b-a$ determines whether $0<\theta<\tfrac{1}{4}\pi$
or $\tfrac{1}{4}\pi<\theta<\half\pi$.  The former corresponds to $a<b$ while the latter
corresponds to $a>b$.  
Likewise, for $c<0$, the sign of $b-a$ determines whether $-\half\pi<\theta< -\tfrac{1}{4}\pi$
or $-\tfrac{1}{4}\pi<\theta<0$.  The former corresponds to $a>b$ while the latter
corresponds to $a<b$.  The borderline cases are likewise determined:
\beqa
&&a=b \quad {\rm and} \quad c\neq  0\quad\Longrightarrow\quad \sin 2\theta=\sgn(c)\,,\nonumber\\[6pt] 
&&a\neq b \quad {\rm and} \quad c= 0\quad\Longrightarrow\quad \cos 2\theta=\sgn(b-a)\,,\nonumber
\eeqa
If $a=b$ and $c=0$, then $A=a\id$, in which case $U$ is arbitrary.
\clearpage

 \section{The singular value decomposition of a complex \texorpdfstring{$2\times 2$}{2x2} matrix}
 \label{sec:SVD}
 \bigskip
 
 For any complex $n\times n$ matrix $M$,
unitary $n\times n$ matrices $L$ and $R$ exist such that
\beq
\label{app:svd}
L^{\T} M R= M_D={\rm diag}(m_1,m_2,\ldots,m_n),
\eeq
where the $m_k$ are real and nonnegative.  This is called the singular
value decomposition of the matrix~$M$.  
A proof of \eq{app:svd} is given
in Appendix D of Ref.~\cite{Dreiner:2008tw} (see also Refs.~\cite{horn,horn2}).
In general, the $m_k$ are
\textit{not} the eigenvalues of $M$.  Rather, the $m_k$ are the
\textit{singular values}
of the general complex matrix $M$, which are
defined to be the nonnegative square roots of the eigenvalues
of either $M^\dagger M$ or $MM^\dagger$ (both yield the same results).

An equivalent definition of the singular values can be
established as follows.  Since $M^\dagger M$ is a 
nonnegative hermitian matrix, its eigenvalues are real and nonnegative and its
eigenvectors, $w_k$, defined by $M^\dagger M w_k = m_k^2 w_k$,
can be chosen to be orthonormal.\footnote{We define
the inner product of two vectors
to be $\ip{v}{w}\equiv v^\dagger w$.}
Consider first the eigenvectors corresponding to the positive
eigenvalues of  $M^\dagger M$.  Then, we
define the vectors $v_k$ such that
$M w_k= m_k v_k^*$.
It follows that $m_k^2 w_k=M^\dagger M w_k= m_k M^\dagger v_k^*$,
which yields: $M^\dagger v_k^*=m_k w_k$.  Note that these equations
also imply that $MM^\dagger v_k^*=m^2_k v_k^*$.
The orthonormality of the $w_k$ implies the
orthonormality of the $v_k^*$ (and hence the $v_k$):
\beq \label{dww}
\delta_{jk}=\ip{w_j}{w_k}=\frac{1}{m_j m_k}
\ip{M^\dagger v^*_j}{M^\dagger v^*_k}=\frac{1}{m_j m_k}
\ip{v^*_j}{MM^\dagger v^*_k}=\frac{m_k}{m_j}\ip{v_j^*}{v_k^*}\,,
\eeq
which yields $\ip{v_j^*}{v_k^*}=\delta_{jk}$.

If $w_i$ is an eigenvector of $M^\dagger M$ with zero eigenvalue, then
$0=w_i^\dagger M^\dagger M w_i=\ip{ M w_i}{M w_i}$, which implies that
$M w_i=0$.
Likewise, if $v_i^*$ is an eigenvector of $MM^\dagger$ with zero
eigenvalue, then $0=v_i^{\T} M M^\dagger v_i^*=\ip{ M^{\T} v_i}{M^{\T}
v_i}^*$, which implies that $M^{\T} v_i=0$. Because the eigenvectors of
$MM^\dagger$
[$M^\dagger M$] can be chosen orthonormal, the eigenvectors
corresponding to the zero eigenvalues of $M$ [$M^{\T}$] can be taken to
be orthonormal.\footnote{The multiplicity of zero eigenvalues of
$M^\dagger M$ [$MM^\dagger$],
which is equal to the number of linearly independent
eigenvectors of $M^\dagger M$ [$MM^\dagger$]
with zero eigenvalue, coincides with the number
of linearly independent eigenvectors of $M$ [$M^{\T}$] with zero
eigenvalue.  Moreover, the number of linearly independent $w_i$
coincides with the number of linearly independent~$v_i$.}
Finally, these eigenvectors are also orthogonal
to the eigenvectors corresponding to the nonzero eigenvalues of
$MM^\dagger$ [$M^\dagger M$].  That is,
\beq \label{ortho2}
\ip{w_j}{w_i}=\frac{1}{m_j}\ip{M^\dagger v_j^*}{w_i}
=\frac{1}{m_j}\ip{v_j^*}{Mw_i} =0\,,
\eeq
and similarly $\ip{v_j}{v_i}=0$, where the index $i$ [$j$] runs over the
eigenvectors corresponding to the zero [nonzero] eigenvalues.
Thus, we can define the
singular values of a general complex matrix $M$ to be the simultaneous
solutions (with real nonnegative $m_k$)
of,\footnote{One can always find a
solution to \eq{singvals} such that the $m_k$ are real and
nonnegative.  Given a solution where $m_k$ is complex,
we simply write $m_k=|m_k|e^{i\theta}$
and redefine $v_k\to v_k e^{i\theta}$ to remove the phase $\theta$.}
\beq
\label{singvals}
Mw_k=m_k v_k^*\,,\qquad\quad v_k^{\T} M =m_k w_k^\dagger\,.
\eeq
The corresponding $v_k$ ($w_k$), normalized to have unit
norm,
are called the left (right) singular vectors of~$M$.
\clearpage

The singular value decomposition of a general $2\times 2$ complex matrix
can be performed fully analytically. The result is more involved than the
standard diagonalization of a $2\times 2$ hermitian matrix by a unitary similarity transformation.
Let us consider the non-diagonal complex matrix,
\beqa
M = \left(\begin{array}{cc}
            a      &  \quad  c  \\
      \tilde{c}    &\quad    b
          \end{array}\right)\,,
\eeqa
where at least one of the two quantities $c$ or $\tilde{c}$ is nonzero. The singular value decomposition of the complex matrix $M$ is
\beq \label{LMR22}
L^{\T} MR=\begin{pmatrix} m_1 & \,\,\, 0 \\ 0 & \,\,\, m_2\end{pmatrix}\,,
\eeq
where $L$ and $R$ are unitary $2\times 2$ matrices and $m_1$, $m_2$ are nonnegative.   Following
Ref.~\cite{murnaghan}, one can parameterize the matrices $L$ and $R$ as
follows,\footnote{Without loss of generality, we have employed the same diagonal phase matrix
$P$ in defining $L$ and $R$.  Had we written $L=U_L P_L$ and $R=U_R P_R$ in \eqs{eq:ulpl}{eq:urpr} with $P_{L,R}\equiv {\rm diag}(e^{-i\alpha_{L,R}}\,,\,e^{-i\beta_{L,R}})$, we would have discovered that only the sums
$\alpha_L +\alpha_R$ and $\beta_L+\beta_R$ are fixed.  Moreover, since \eq{LMR22} is unchanged under $\alpha\to\alpha+\pi$ or $\beta\to\beta+\pi$, one can fix the range of $\alpha$ and $\beta$ as specified below \eq{eq:urpr}.}
%
%
\beqa
&& L = U_L P
     = \left(\begin{array}{cc}
         \cos\theta_L   &    e^{i \phi_L} \sin\theta_L  \\
     -e^{-i\phi_L} \sin\theta_L & \cos\theta_L
             \end{array}\right) \,
       \left(\begin{array}{cc}
          e^{-i\alpha}  &  0 \\
          0  & e^{-i\beta}
             \end{array}\right)\,,\label{eq:ulpl} \\[10pt]
&& R = U_R P
     = \left(\begin{array}{cc}
         \cos\theta_R   &    e^{i \phi_R} \sin\theta_R  \\
     -e^{-i\phi_R} \sin\theta_R & \cos\theta_R
       \end{array}\right)\,
       \left(\begin{array}{cc}
          e^{-i\alpha}  &  0 \\
          0  & e^{-i\beta}
             \end{array}\right)\,,\label{eq:urpr}
\eeqa
where $0\leq \theta_{L,R} \leq \half\pi$, $0\leq \alpha,\beta< \pi$, and $0\leq\phi_L,\phi_R<2\pi$.

The singular values $m_{1,2}$ of the matrix $M$ 
can be determined by taking the positive square root of the nonnegative
eigenvalues, $m^2_{1,2}$, of the hermitian matrix $M^\dagger M$,
\beq \label{msquared}
m^2_{1,2}= \tfrac{1}{2}\bigl[|a|^2+|b|^2+|c|^2+|\tilde{c}|^2 \mp \Delta\bigr]\,,
\eeq
in a convention where $0\leq m_1\leq m_2$ (i.e., $\Delta\geq 0$), with
\beqa
\Delta &\equiv &\bigl[(|a|^2-|b|^2-|c|^2+|\tilde{c}|^2)^2
                    +4|a^* c + b \tilde{c}^*|^2\bigr]^{1/2}\nonumber \\[6pt]
                    &=& \bigl[(|a|^2+|b|^2+|c|^2+|\tilde{c}|^2)^2
                    -4|a b - c \tilde{c}|^2\bigr]^{1/2}\,. \label{Deltasqlong}
\eeqa
It follows that
\beq \label{sumprodDel}
m_1^2+m_2^2=|a|^2+|b|^2+|c|^2+|\tilde{c}|^2\,,\qquad\quad \Delta=m_2^2-m_1^2\,.
\eeq
Moreover, by taking the determinant of \eq{LMR22}, it follows that
\beq \label{emonetwo}
m_1 m_2=(ab-c\tilde{c})e^{-2i(\alpha+\beta)}\,.
\eeq
Note that $m_1=m_2$ if and only if $|a|=|b|$, $|c|=|\tilde{c}|$ and $a^* c + b \tilde{c}^*=0$
are satisfied.  

We first assume that $m_1\neq m_2$.
Using the results of Section~\ref{diaghermitian} enables us to compute the rotation angles, $\theta_{L,R}$,
and the phases, $e^{i\phi_{L,R}}$,  by diagonalizing
$M^\dagger M$ and $M^*M^{\T}$ with a diagonalizing matrix $R$ and $L$, respectively.
Explicitly, we have
\beq \label{MdaggerM}
M^\dagger M=\begin{pmatrix}  |a|^2+|\tilde{c}|^2 & \quad a^* c+b\tilde{c}^* \\  ac^*+ b^*\tilde{c} & \quad |b|^2+|c|^2\end{pmatrix}\,,
\eeq
and $M^*M^{\T}$ is obtained form $M^\dagger M$ by interchanging $c$ and $\tilde{c}$.  Applying \eqs{chiresult}{sctheta} to the diagonalization of 
$M^\dagger M$ and $M^*M^{\T}$ then yields,
\beq \label{eq:ctheta_LR}
\cos\theta_{R,L} = \sqrt{\frac{\Delta + |b|^2-|a|^2
                           \pm |c|^2 \mp |\tilde{c}|^2}{2\Delta}}\,, \qquad\quad
                             \sin\theta_{R,L} = \sqrt{\frac{\Delta - |b|^2+|a|^2
                             \mp |c|^2\pm |\tilde{c}|^2}{2\Delta}}\,,
\eeq
and
\beq
\label{eq:phi_LR}
e^{i\phi_R} = \frac{a^* c + b \tilde{c}^*}{|a^* c + b \tilde{c}^*|}\,,\qquad\quad
e^{i\phi_L} = \frac{a^* \tilde{c} + b c^*}{|a^* \tilde{c} + b c^*|}\,.
\eeq
For completeness, we note that the denominators in \eq{eq:phi_LR} can be written in another form by employing the following results [which are a consequence of \eq{Deltasqlong}],
\beqa 
|a^* c+b\tilde{c}^*|&=&\half\sqrt{\Delta^2-(|b|^2-|a|^2+|c|^2-|\tilde{c}|^2)^2}\,, \label{modacbc} \\[6pt]
|a^* \tilde{c}+bc^*|&=&\half\sqrt{\Delta^2-(|b|^2-|a|^2-|c|^2+|\tilde{c}|^2)^2}\,.
\eeqa

The final step of the computation is to determine the angles $\alpha$ and
$\beta$.  To perform this task, we first rewrite \eq{LMR22} as,
\beq \label{MURULD}
MU_R=U_L^*\begin{pmatrix} m_1 e^{2i\alpha} & \, 0 \\ 0 & \,m_2 e^{2i\beta}\end{pmatrix}\,,
\eeq
where we have made use of \eqs{eq:ulpl}{eq:urpr}.  Setting the diagonal elements of the left hand side and the right hand side of \eq{MURULD} equal, we end up with the following two equations,
\beqa
m_1 \cos\theta_L e^{2i\alpha}&=& a \cos\theta_R-c \,e^{-i\phi_R}\sin\theta_R\,, \label{m1e} \\[3pt]
m_2 \cos\theta_L e^{2i\beta}&=& b \cos\theta_R+\tilde{c}\, e^{i\phi_R}\sin\theta_R\,. \label{m2e}
\eeqa
Next, we multiply both \eqs{m1e}{m2e} by $\Delta\,\cos\theta_R$.   Employing \eqst{eq:ctheta_LR}{eq:phi_LR} on the right hand sides of the two resulting equations then yields,
\beqa
\Delta\, m_1 \cos\theta_L \cos\theta_R  e^{2i\alpha}&=& \half a\bigl(\Delta+|b|^2-|a|^2+|c|^2-|\tilde{c}|^2\bigr) \nonumber \\
&& \qquad  -\frac{c(ac^*+b^*\tilde{c})}{2|a^* c+b\tilde{c}^*|}\sqrt{\Delta^2-(|b|^2-|a|^2+|c|^2-|\tilde{c}|^2)^2}\,,  \label{m1ccD}\\[3pt]
\Delta\, m_2 \cos\theta_L \cos\theta_R e^{2i\beta}&=& \half b\bigl(\Delta+|b|^2-|a|^2+|c|^2-|\tilde{c}|^2\bigr) \nonumber \\
&& \qquad   -\frac{\tilde{c}(a^*c+b\tilde{c}^*)}{2|a^* c+b\tilde{c}^*|}\sqrt{\Delta^2-(|b|^2-|a|^2+|c|^2-|\tilde{c}|^2)^2}\,. \label{m2ccD}
\eeqa
We can simplify \eqs{m1ccD}{m2ccD} further by making use of \eq{modacbc}.  The end result is,
\beqa 
\Delta\,m_1 \cos\theta_L \cos\theta_R  e^{2i\alpha}&=& \half a\bigl(\Delta+|b|^2-|a|^2-|c|^2-|\tilde{c}|^2\bigr) -b^* c\tilde{c}\,, \label{m1LR} \\[3pt]
\Delta \,m_2 \cos\theta_L \cos\theta_R e^{2i\beta}&=& \half b\bigl(\Delta+|b|^2-|a|^2+|c|^2+|\tilde{c}|^2\bigr)  +a^* c\tilde{c}\,.\label{m2LR}
\eeqa
Using \eq{msquared}, it is convenient to eliminate $\Delta$ in favor of $m_1^2$ and $m_2^2$ on the right hand side of \eqs{m1LR}{m2LR}.
It then immediately follows that,
\beqa
\alpha &=& \half\arg\bigl\{a\bigl(|b|^2-m_1^2\bigr)-b^*c\tilde{c}\bigr\}
\label{argalph}\,,\\[6pt]
\beta &=& \half\arg\bigl\{b\bigl(m_2^2-|a|^2\bigr)+a^*c\tilde{c}\bigr\}\,.
\label{argbet}
\eeqa

A useful identity can now be derived that exhibits a simple relation between the angles $\theta_L$ and $\theta_R$.  
First, we make use \eq{eq:ctheta_LR} to obtain,
\beqa
\cos 2\theta_L&=&\frac{|b|^2-|a|^2-|c|^2+|\tilde{c}|^2}{\Delta}\,,\qquad\quad\,\cos 2\theta_R=\frac{|b|^2-|a|^2+|c|^2-|\tilde{c}|^2}{\Delta}\,,\label{svdc2LR}\\
\sin 2\theta_L&=&\frac{|a^*\tilde{c}+bc^*|}{\Delta}\,,\qquad\qquad\qquad\qquad \sin 2\theta_R=\frac{|a^*c+b\tilde{c}^*|}{\Delta}\,.\label{svds2LR}
\eeqa
Next, we note two different trigonometric identities for the tangent function to obtain,
\beqa
\tan\theta_L&=&\frac{1-\cos 2\theta_L}{\sin 2\theta_L} =\frac{m_2^2-m_1^2-|b|^2+|a|^2+|c|^2-|\tilde{c}|^2}{2|a^*\tilde{c}+bc^*|}=\frac{|a|^2+|c|^2-m_1^2}{|a^*\tilde{c}+bc^*|}\,,\label{tantL}\\[6pt]
\tan\theta_R&=&\frac{\sin 2\theta_R}{1+\cos 2\theta_R}=\frac{2|a^*c+b\tilde{c}^*|}{m_2^2-m_1^2+|b|^2-|a|^2+|c|^2-|\tilde{c}|^2}=\frac{|a^*c+b\tilde{c}^*|}{m_2^2-|a|^2-|\tilde{c}|^2}\,,\label{tantR}
\eeqa
where we have made use of \eqss{sumprodDel}{svdc2LR}{svds2LR}.  It then follows that
\beq \label{tLtRc}
\frac{\tan\theta_L}{\tan\theta_R}=\frac{(|a|^2+|c|^2-m_1^2)(m_2^2-|a|^2-|\tilde{c}|^2)}{|(a^*\tilde{c}+bc^*)(a^*c+b\tilde{c}^*)|}\,.
\eeq
The numerator of \eq{tLtRc} can be simplified with a little help from \eqs{sumprodDel}{emonetwo} as follows,
\beqa
 (|a|^2+|c|^2-m_1^2)(m_2^2-|a|^2-|\tilde{c}|^2)&=&|a|^2(m_1^2+m_2^2)+|c|^2 m_2^2-|\tilde{c}|^2m_1^2-m_1^2m_2^2 \nonumber \\
 && \qquad\qquad  -(|a|^2+|c|^2)(|a|^2+|\tilde{c}|^2) \nonumber \\[6pt]
&=& |a|^2(|a|^2+|b|^2+|c|^2+|\tilde{c}|^2)+|c|^2 m_2^2+|\tilde{c}|^2 m_1^2 \nonumber \\
&& \qquad\qquad -|ab-c\tilde{c}|^2-(|a|^2+|c|^2)(|a|^2+|\tilde{c}|^2) \nonumber \\[6pt]
&=&  |c|^2 m_2^2+|\tilde{c}|^2 m_1^2+(ab-c\tilde{c})c^*\tilde{c}^*+(a^*b^*-c^*\tilde{c}^*)c\tilde{c} \nonumber \\[6pt]
&=& (c m_2 e^{-i(\alpha+\beta)}+\tilde{c}^* m_1 e^{i(\alpha+\beta)})(c^*m_2 e^{i(\alpha+\beta)}+\tilde{c}m_1 e^{-i(\alpha+\beta)})\,.\nonumber \\
&& \phantom{line}
\eeqa
Likewise, the denominator of \eq{tLtRc} can be simplified as follows,
\beqa
|(a^*\tilde{c}+bc^*)(a^*c+b\tilde{c}^*)|&=&|(a\tilde{c}^*+b^*c)(a^*c+b\tilde{c}^*)|
= |c\tilde{c}^*(|a|^2+|b|^2)+ab\tilde{c}^{*\,2}+a^* b^*c^2| \nonumber \\[6pt]
&=& |c\tilde{c}^*(|a|^2+|b|^2+|c|^2+|\tilde{c}|^2)+(ab-c\tilde{c})c^{*\,2}+(a^* b^*-c^*\tilde{c}^*)c^2| \nonumber \\[6pt]
&=&|c\tilde{c}^*(m_1^2+m_2^2)+m_1 m_2(c^{*\,2} e^{2i(\alpha+\beta)}+c^2 e^{-2i(\alpha+\beta)})| \nonumber \\[6pt]
&=& |(cm_2 e^{-i(\alpha+\beta)}+\tilde{c}^* m_1 e^{i(\alpha+\beta)})(\tilde{c}^* m_2 e^{i(\alpha+\beta)}+cm_1 e^{-i(\alpha+\beta)})|\,.
\eeqa
Hence, we end up with a remarkably simple result,
\beq \label{amusingidc}
\frac{\tan\theta_L}{\tan\theta_R}=\left|\frac{c^*m_2 e^{i(\alpha+\beta)}+\tilde{c}m_1 e^{-i(\alpha+\beta)}}{\tilde{c}^* m_2 e^{i(\alpha+\beta)}+cm_1 e^{-i(\alpha+\beta)}}\right|\,.
\eeq
If $m_1\neq 0$, then one can employ \eq{emonetwo} to obtain an alternate form for \eq{amusingidc},
\beq
\frac{\tan\theta_L}{\tan\theta_R}=\left|\frac{c^* (ab-c\tilde{c})+\tilde{c}m_1^2}{\tilde{c}^*(ab-c\tilde{c})+cm_1^2}\right|\,.
\eeq

The case of $m_1=0$ is noteworthy.  This special case arises when $\det M =ab - c\tilde{c}=0$, which implies that there is one singular
value that is equal to zero.   In particular, it then follows that
$\Delta=|a|^2+|b|^2+|c|^2+|\tilde{c}|^2$ [cf.~\eq{Deltasqlong}] and 
\beq \label{mtwospecial}
m_2^2=\Tr(M^\dagger M)=|a|^2+|b|^2+|c|^2+|\tilde{c}|^2\,.   
\eeq
\Eqss{tantL}{tantR}{mtwospecial} then yield,\footnote{If either $c=0$ or $\tilde{c}=0$ then $ab=0$, in which case one should discard any fractions appearing in \eqs{emonezero1}{emonezero2} that are of indeterminate form.}
\beq \label{emonezero1}
\tan\theta_L=\left|\frac{c}{b}\right|=\left|\frac{a}{\tilde{c}}\right|\,,\qquad\qquad \tan\theta_R=\left|\frac{a}{c}\right|=\left|\frac{\tilde{c}}{b}\right|\,,
\eeq
after using $c\tilde{c}=ab$, and
\beq \label{emonezero2}
\phi_L=\arg(b/c)=\arg(\tilde{c}/a)\,,\qquad \phi_R=\arg(c/a)=\arg(b/\tilde{c})\,,\qquad  \beta=\half\arg b\,.
\eeq
As expected the angle $\alpha$ is undefined when $m_1=0$ [cf.~\eqs{m1LR}{argalph}].

Finally, we treat the case of degenerate nonzero singular
values, i.e.~$m\equiv m_1=m_2\neq 0$.   As previously noted below \eq{Deltasqlong},
degenerate singular values exist if and only if 
\beq \label{degenconds}
|a|=|b|\,,\, |c|=|\tilde{c}|\,,\, \text{and $a^* c = - b \tilde{c}^*$}.
\eeq  
Note that \eq{degenconds} also implies that $a^*\tilde{c}=-bc^*$.
It then follows from \eq{MdaggerM} that
\beq
M^\dagger M = m^2\mathds{1}_{2\times 2}\,,
\eeq
where the degenerate singular value is
\beq \label{msingvalue}
m=\sqrt{|a|^2+|c|^2}\,.
\eeq
Hence, the diagonalization equation, $R^{-1}M^\dagger MR=m^2\mathds{1}_{2\times 2}$, is satisfied for any unitary matrix~$R$.  However, this does not necessarily mean that
an arbitrary unitary matrix $R$ is a solution to \eq{LMR22}.  In the analysis given below, we shall see that in the case of degenerate singular values,
$\alpha+\beta$ is fixed by the matrix $M$, whereas the remaining parameters that define the matrix $R$ exhibited in \eq{eq:urpr} can be taken as arbitrary.

Given the unitary matrix $R$, one can use \eq{LMR22} to determine the matrix elements of the unitary matrix $L$.  
Using \eqs{eq:ulpl}{eq:urpr}, it follows that
\beq \label{ULT}
U_L^{\T}=m\begin{pmatrix} e^{2i\alpha} & \,\,\, 0 \\ 0 & \,\,\, e^{2i\beta}\end{pmatrix} U_R^\dagger M^{-1}\,.
\eeq
In light of \eqs{degenconds}{msingvalue},
\beq
\det M=ab-c\tilde{c}=-\frac{cm^2}{\tilde{c}^*}\,.
\eeq
Evaluating the left and right hand sides of \eq{ULT} yields,
\beqa
\cos\theta_L &=& -\frac{\tilde{c}^*}{mc}e^{2i\alpha}\bigl(b\cos\theta_R+\tilde{c}e^{i\phi_R}\sin\theta_R\bigr)=-\frac{\tilde{c}^*}{mc}e^{2i\beta}\bigl(a\cos\theta_R-ce^{-i\phi_R}\sin\theta_R\bigr)\,, \label{CL1}
\\[6pt]
e^{i\phi_L}\sin\theta_L &=& \frac{\tilde{c}^*}{mc}e^{2i\beta}\bigl(\tilde{c}\cos\theta_R-be^{-i\phi_R}\sin\theta_R\bigr)=-\frac{\tilde{c}}{mc^*}e^{-2i\alpha}\bigl(c^*\cos\theta_R+a^*e^{-i\phi_R}\sin\theta_R\bigr)\,.\label{SL1}
\eeqa
We can rewrite the first part of \eq{CL1} as follows,
\beqa
m\cos\theta_L &=& -e^{-2i\alpha}\left(\frac{b^*\tilde{c}}{c^*}b^*\cos\theta_R+\frac{|\tilde{c}|^2}{c^*}e^{-i\phi_R}\sin\theta_R\right) \nonumber \\[6pt]
&=&  e^{-2i\alpha}\bigl(a\cos\theta_R-c e^{-i\phi_R}\sin\theta_R\bigr)\,,
\eeqa
after complex conjugating and making use of \eq{degenconds}.   A similar manipulation (without the complex conjugation) can be performed on the last term of \eq{SL1}.
The end result is
\beqa
m\cos\theta_L &=&  e^{-2i\alpha}\bigl(a\cos\theta_R-c e^{-i\phi_R}\sin\theta_R\bigr) = -\frac{\tilde{c}^*}{c}e^{2i\beta}\bigl(a\cos\theta_R-ce^{-i\phi_R}\sin\theta_R\bigr)\,,  \label{CL2} \\[6pt]
m e^{i\phi_L}\sin\theta_L &=& \frac{\tilde{c}^*}{c}e^{2i\beta}\bigl(\tilde{c}\cos\theta_R-be^{-i\phi_R}\sin\theta_R\bigr)=-e^{-2i\alpha}\bigl(\tilde{c}\cos\theta_R-be^{-i\phi_R}
\sin\theta_R\bigr)\,.\label{SL2}
\eeqa
Since both \eqs{CL2}{SL2} cannot simultaneously vanish, it follows that 
\beq \label{cctil}
e^{2i(\alpha+\beta)}=-\frac{c}{\tilde{c}^*}\,.
\eeq
We conclude that if $\theta_R$, $\phi_R$ and $\alpha-\beta$ are taken to be arbitrary parameters, then $\theta_L$ and $\phi_L$ are fixed by \eqs{CL2}{SL2} and $\alpha+\beta$ is fixed by \eq{cctil}.  In Appendix A, we show how to employ \eqs{CL2}{SL2} to construct explicit examples of the singular decomposition of a $2\times 2$ complex matrix $M$ that possesses degenerate singular values.

For a simple example of the degenerate case, consider the singular value decomposition of the matrix,
\beq
M=\begin{pmatrix} 0 & \,\,\, 1 \\ 1 & \,\,\, 0\end{pmatrix}\,.
\eeq
Setting $a=b=0$ and $c=\tilde{c}=m=1$ in \eqst{CL2}{cctil}, it then follows that
\beq
\cos\theta_L=e^{i(2\beta-\phi_R)}\sin\theta_R\,,\qquad\quad  \sin\theta_L=e^{i(2\beta-\phi_L)}\cos\theta_R\,,\qquad \quad e^{-2i\alpha}=-e^{2i\beta}\,.
\eeq
Hence, we conclude that $\phi_L=\phi_R\equiv\phi$, $\theta_L=\half\pi-\theta_R$,
$\beta=\half\phi$ and $\alpha=-\half(\phi\pm\pi)$.   Plugging these values into \eqs{eq:ulpl}{eq:urpr}, we obtain
\beq \label{LTR}
L=\begin{pmatrix} \pm ie^{i\phi/2}\sin\theta_R & \,\,\,  e^{i\phi/2}\cos\theta_R \\  \mp ie^{-i\phi/2}\cos\theta_R & \,\,\, e^{-i\phi/2}\sin\theta_R\end{pmatrix}\,, \qquad\quad
R=\begin{pmatrix} \pm ie^{i\phi/2}\cos\theta_R & \,\,\, e^{i\phi/2}\sin\theta_R \\ \mp ie^{-i\phi/2}\sin\theta_R & \,\,\, e^{-i\phi/2}\cos\theta_R\end{pmatrix}\,.
\eeq
One can check that $L^{\T}MR=\mathds{1}_{2\times 2}$.  Thus, we have found a family of singular value decompositions of $M$ that depend on two parameters $\theta_R$ and $\phi$.
This does not exhaust all possible singular value decompositions of $M$, since one is always free to multiply $R$ on the right by $Q\,{\rm diag}( e^{-i\chi_1}\,,\,e^{-i\chi_2})$ and multiply $L$ on the right by $Q\,{\rm diag}( e^{i\chi_1}\,,\,e^{i\chi_2})$, where $Q$ is an arbitrary real orthogonal $2\times 2$ matrix and $0\leq\chi_i<2\pi$. 

We shall now exhibit two different singular value decompositions of $M$.   First, if we choose the lower signs in \eq{LTR}, with $\theta_R=\phi=\half\pi$, $Q=\mathds{1}_{2\times 2}$ and $\chi_1=\chi_2=\tfrac14\pi$, then it follows that  
\beq
L=\mathds{1}_{2\times 2}\,,\qquad\quad R=\begin{pmatrix} 0 & \,\,\, 1 \\ 1 & \,\,\, 0\end{pmatrix}\,.
\eeq
Second, choosing the upper signs in \eq{LTR} with $\theta_R=\tfrac14\pi$, $\phi=\chi_1=\chi_2=0$ and $Q=\mathds{1}_{2\times 2}$  yields,
\beq
L=R=\frac{1}{\sqrt{2}}\begin{pmatrix} \phm i & \,\,\, 1 \\ -i & \,\,\, 1\end{pmatrix}\,.
\eeq
A singular value decomposition with $L=R$
corresponds to an Autonne-Takagi factorization of a complex symmetric matrix $M$.  This is the subject of the Section~\ref{sec:takagi}.

 \section{The singular value decomposition of a real \texorpdfstring{$2\times 2$}{2x2} matrix over the space of real matrices}
 \label{sec:realSVD}
 \bigskip
 
 For any real $n\times n$ matrix $M$,
real orthogonal $n\times n$ matrices $L$ and $R$ exist such that
\beq
\label{real:svd}
L^{\T} M R= M_D={\rm diag}(m_1,m_2,\ldots,m_n),
\eeq
where the $m_k$ are real and nonnegative.  This corresponds to the \textit{real} singular value decomposition of $M$, which is restricted to the space of real matrices.  A separate treatment independent of the one presented in Section~\ref{sec:SVD} is warranted.
As in the complex case treated in Section~\ref{sec:SVD}, the $m_k$ are
\textit{not} the eigenvalues of~$M$.  Rather, the $m_k$ are the
singular values
of a real matrix $M$, which are
defined to be the nonnegative square roots of the eigenvalues of either
$M^{\T} M$ or $MM^{\T}$ (both yield the same results).

An equivalent definition of the singular values can be
established as follows.  Since $M^{\T} M$ is a nonnegative real symmetric
matrix, its eigenvalues are real and nonnegative and its
eigenvectors, $w_k$, defined by $M^{\T} M w_k = m_k^2 w_k$,
can be chosen to be real and orthonormal.
First, consider the eigenvectors of  $M^{\T} M$ corresponding to the positive
eigenvalues, $m_k\neq 0$.  We then
define the vectors $v_k$ such that
$M w_k= m_k v_k$.
It follows that $m_k^2 w_k=M^{\T} M w_k= m_k M^{\T}v_k$,
which yields $M^{\T} v_k=m_k w_k$.  Note that these equations
also imply that $MM^{\T} v_k=m^2_k v_k$.
The orthonormality of the $w_k$ implies the
orthonormality of the $v_k$,
\beq \label{dwwr}
\delta_{jk}=\ip{w_j}{w_k}=\frac{1}{m_j m_k}
\ip{M^{\T} v_j}{M^{\T} v_k}=\frac{1}{m_j m_k}
\ip{v_j}{MM^{\T}v_k}=\frac{m_k}{m_j}\ip{v_j}{v_k}\,,
\eeq
which yields $\ip{v_j}{v_k}=\delta_{jk}$.

Second, if $w_i$ is an eigenvector of $M^{\T} M$ with zero eigenvalue $m_i=0$, then it follows that
$0=w_i M^{\T} M w_i=\ip{ M w_i}{M w_i}$, which implies that
$M w_i=0$.
Likewise, if $v_i$ is an eigenvector of $MM^{\T}$ with zero
eigenvalue, then $0=v_i^{\T} M M^{\T} v_i=\ip{ M^{\T} v_i}{M^{\T}
v_i}$, which implies that \mbox{$M^{\T} v_i=0$}. Because the eigenvectors of
$MM^{\T}$
[$M^{\T} M$] can be chosen orthonormal, the eigenvectors
corresponding to the zero eigenvalues of $M$ [$M^{\T}$] can be taken to
be orthonormal.
Finally, these eigenvectors are also orthogonal
to the eigenvectors corresponding to the nonzero eigenvalues of
$MM^{\T}$ [$M^{\T} M$].  That is,
\beq \label{ortho2r}
\ip{w_j}{w_i}=\frac{1}{m_j}\ip{M^{\T} v_j}{w_i}
=\frac{1}{m_j}\ip{v_j}{Mw_i} =0\,,
\eeq
and similarly $\ip{v_j}{v_i}=0$, where the index $i$ [$j$] runs over the
eigenvectors corresponding to the zero [nonzero] eigenvalues.
Thus, we can define the
singular values of a real matrix $M$ to be the simultaneous
solutions (with real nonnegative $m_k$)
of,\footnote{One can always find a
solution to \eq{singvals} such that the $m_k$ are real and
nonnegative.  Given a solution where $m_k$ is complex,
we simply write $m_k=|m_k|e^{i\theta}$
and redefine $v_k\to v_k e^{i\theta}$ to remove the phase $\theta$.}
\beq
\label{singvalsr}
Mw_k=m_k v_k\,,\qquad\quad v_k^{\T} M =m_k w_k^{\T}\,.
\eeq
The corresponding $v_k$ ($w_k$), normalized to have unit
norm,
are called the left (right) singular vectors of~$M$.

The real singular value decomposition of a general $2\times 2$ real matrix
can be performed fully analytically.
Let us consider the non-diagonal real matrix,
\beqa
M = \left(\begin{array}{cc}
            a      &  \quad  c  \\
      \tilde{c}    &\quad    b
          \end{array}\right)\,,
\eeqa
where at least one of the two quantities $c$ or $\tilde{c}$ is nonzero.  The real singular value decomposition of the real matrix $M$ is
\beq \label{LMR2r}
L^{\T} MR=\begin{pmatrix} m_1 & \,\,\, 0 \\ 0 & \,\,\, m_2\end{pmatrix}\,,
\eeq
where $L$ and $R$ are real $2\times 2$ orthogonal matrices and $m_1$, $m_2$ are nonnegative.
In general, one can parameterize
$L$ and $R$ in \eq{LMR2r} by %
\beq
 L 
     = \left(\begin{array}{cc}
        \phm \cos\theta_L   &     \sin\theta_L  \\
     - \sin\theta_L &  \cos\theta_L
             \end{array}\right)\begin{pmatrix} 1 & \,\,\,0 \\ 0 & \,\,\, \varepsilon_L\end{pmatrix}\,,\qquad\quad
 R 
     = \left(\begin{array}{cc}
        \phm \cos\theta_R   &   \sin\theta_R  \\
     - \sin\theta_R & \cos\theta_R
       \end{array}\right)\begin{pmatrix} 1 & \,\,\, 0 \\ 0 & \,\,\, \varepsilon_R\end{pmatrix}\,,\label{eq:OLR}
\eeq
where $-\half\pi<\theta_{L,R} \leq \half\pi$, and $\varepsilon_{L,R}=\pm 1$.
Note that $\det L=\varepsilon_L$ and
$\det R=\varepsilon_R$, which implies that $\varepsilon_L\varepsilon_R\det M=m_1 m_2$.  Since $m_1$, $m_2\geq 0$, it follows that $\sgn(\det M)=\varepsilon_L\varepsilon_R$.   Thus, only the product of $\varepsilon_L$ and $\varepsilon_R$ is fixed by \eq{LMR2r}.

The parameterization of $L$ and $R$ given in \eq{eq:OLR} is related to that of
\eqs{eq:ulpl}{eq:urpr} as follows.   When $M$ is a real matrix, the quantities $e^{i\phi_L}=\sgn(a\tilde{c}+bc)$ and
$e^{i\phi_R}=\sgn(ac+b\tilde{c})$.
Hence, we can set $\phi_L=\phi_R=0$ and redefine $\theta_L\to \theta_L\sgn(a\tilde{c}+bc)$ and $\theta_R\to \theta_R\sgn(ac+b\tilde{c})$, thereby extending the range of these angular variables to
$-\half\pi<\theta_{L,R}\leq\half\pi$ as indicated above.  Finally, it is convenient to replace the phase matrix $P$ in 
\eqs{eq:ulpl}{eq:urpr} with ${\rm diag}(1,\varepsilon_L)$ and  ${\rm diag}(1,\varepsilon_R)$, respectively, so that the matrices $L$ and $R$ are real orthogonal matrices
 (rather than the more general unitary matrices).

The singular values $m_{1,2}$ of the matrix $M$ 
can be determined by taking the positive square root of the nonnegative
eigenvalues, $m^2_{1,2}$, of the real orthogonal matrix $M^{\T}M$,
\beq \label{msquaredreal}
m^2_{1,2}= \tfrac{1}{2}\bigl[a^2+b^2+c^2+\tilde{c}^2 \mp \Delta\bigr]\,,
\eeq
in a convention where $0\leq m_1\leq m_2$ (i.e., $\Delta\geq 0$), with
\beqa
\Delta &\equiv &\bigl[(a^2-b^2-c^2+\tilde{c}^2)^2
                    +4(ac+b\tilde{c})^2\bigr]^{1/2}\nonumber \\[6pt]
                    &=& \bigl[(a^2+b^2+c^2+\tilde{c}^2)^2
                    -4(a b - c \tilde{c})^2\bigr]^{1/2}\,. \label{Deltasqlong2}
\eeqa
Note that
\beq \label{sumprod}
m_1^2+m_2^2=a^2+b^2+c^2+\tilde{c}^2\,,\qquad\quad m_1 m_2=\varepsilon_L\varepsilon_R(ab-c\tilde{c})\,.
\eeq
Moreover, $m_1=m_2$ if and only if $a=\pm b$ and $c=\mp\tilde{c}$ (where the signs are correlated as indicated), which imply that $ac + b \tilde{c}=0$ and $\Delta=0$.  

We first assume that $m_1\neq m_2$.  Then, if we rewrite \eq{LMR2r} in the form $MR=LM_D$, where $M_D\equiv{\rm diag}(m_1\,,\,m_2)$, then we immediately obtain,
\beqa
m_1 \cos\theta_L&=&a\cos\theta_R-c\sin\theta_R\,,\qquad\quad \varepsilon_L\varepsilon_R m_2\sin\theta_L=a\sin\theta_R+c\cos\theta_R\,,\label{monemtwo}\\
m_1\sin\theta_L&=& b\sin\theta_R-\tilde{c}\cos\theta_R\,,\qquad\quad\,
 \varepsilon_L\varepsilon_R  m_2\cos\theta_L=\tilde{c}\sin\theta_R+b\cos\theta_R\,.\label{monemtwo2}
 \eeqa
 It follows that 
\beq \label{mcs}
m_1^2\cos^2\theta_L+m_2^2\sin^2\theta_L=a^2+c^2\,,\qquad\quad 
m_1^2\sin^2\theta_L+m_2^2\cos^2\theta_L=b^2+\tilde{c}^2\,.
\eeq
Subtracting these two equations, and employing \eq{Deltasqlong2} yields,
\beq \label{c2LR}
\cos 2\theta_L=\frac{b^2-a^2-c^2+\tilde{c}^2}{\Delta}\,,\qquad\quad 
\cos 2\theta_R=\frac{b^2-a^2+c^2-\tilde{c}^2}{\Delta}\,.
\eeq
In obtaining $\cos 2\theta_R$, it is sufficient to note that \eqst{monemtwo}{mcs} are valid under the interchange of $c\leftrightarrow\tilde{c}$ and the interchange of the subscripts $L\leftrightarrow R$.\footnote{One can verify this by rewriting \eq{LMR2r} in the form $L^{\T}M=M_DR^{\T}$, which yields equations of the form given by \eqs{monemtwo}{monemtwo2} with $c\leftrightarrow\tilde{c}$ and the interchange of the subscripts $L\leftrightarrow R$.  Note that $\Delta$ and hence $m_{1,2}^2$ are unaffected by these interchanges.\label{fneleven}}

We can also use \eqs{monemtwo}{monemtwo2} to obtain,
\beqa
m_1^2\cos\theta_L\sin\theta_L=(a\cos\theta_R-c\sin\theta_R)(b\sin\theta_R-\tilde{c}\cos\theta_R)\,,\\
m_2^2\cos\theta_L\sin\theta_L=(a\sin\theta_R+c\cos\theta_R)(\tilde{c}\sin\theta_R+b\cos\theta_R)\,.
\eeqa
Subtracting these two equations yields
\beq \label{s2LR}
\sin 2\theta_L=\frac{2(a\tilde{c}+bc)}{\Delta}\,,\qquad\quad \sin 2\theta_R=\frac{2(ac+b\tilde{c})}{\Delta}\,,
\eeq
after again noting the symmetry under $c\to\tilde{c}$ and the interchange of the subscripts $L\leftrightarrow R$.

Thus, employing \eqs{c2LR}{s2LR}, we have succeeded in uniquely determining the angles $\theta_L$ and $\theta_R$ (where $-\half\pi<\theta_{L,R}\leq\half\pi$).   As noted below \eq{eq:OLR}, the individual signs $\varepsilon_L$ and $\varepsilon_R$ are not separately fixed (implying that one is free to set one of these two signs to $+1$);  only the product $\varepsilon_L\varepsilon_R =\sgn(\det M)$ is determined by the singular value decomposition of $M$.

A useful identity can now be derived that exhibits a simple relation between the angles $\theta_L$ and $\theta_R$.  First, we note two different trigonometric identities for the tangent function,
\beqa
\tan\theta_L&=&\frac{1-\cos 2\theta_L}{\sin 2\theta_L} =\frac{m_2^2-m_1^2-b^2+a^2+c^2-\tilde{c}^2}{2(a\tilde{c}+bc)}=\frac{a^2+c^2-m_1^2}{a\tilde{c}+bc}\,, \label{tanid1} \\[6pt]
\tan\theta_R&=&\frac{\sin 2\theta_R}{1+\cos 2\theta_R}=\frac{2(ac+b\tilde{c})}{m_2^2-m_1^2+b^2-a^2+c^2-\tilde{c}^2}=\frac{ac+b\tilde{c}}{m_2^2-a^2-\tilde{c}^2}\,, \label{tanid2}
\eeqa
where we have made use of \eqss{sumprod}{c2LR}{s2LR}.  It then follows that
\beq \label{tLtR}
\frac{\tan\theta_L}{\tan\theta_R}=\frac{(a^2+c^2-m_1^2)(m_2^2-a^2-\tilde{c}^2)}{(a\tilde{c}+bc)(ac+b\tilde{c})}\,.
\eeq
The numerator of \eq{tLtR} can be simplified with a little help from \eq{sumprod} as follows,
\beqa
 (a^2+c^2-m_1^2)(m_2^2-a^2-\tilde{c}^2)&=&a^2(m_1^2+m_2^2)+c^2 m_2^2-\tilde{c}^2m_1^2-(a^2+c^2)(a^2+\tilde{c}^2)-m_1^2m_2^2 \nonumber \\[6pt]
&=& a^2(a^2+b^2+c^2+\tilde{c}^2)-(a^2+c^2)(a^2+\tilde{c}^2) \nonumber \\
&& \qquad\qquad +c^2 m_2^2+\tilde{c}^2 m_1^2-(ab-c\tilde{c})^2 \nonumber \\[6pt]
&=&  c^2 m_2^2+\tilde{c}^2 m_1^2+2(ab-c\tilde{c})c\tilde{c} =(cm_2+\varepsilon_L\varepsilon_R\tilde{c}m_1)^2\,.
\eeqa
Likewise, the denominator of \eq{tLtR} can be simplified as follows,
\beqa
(a\tilde{c}+bc)(ac+b\tilde{c})
&=& (ab-c\tilde{c})(c^2+\tilde{c}^2)+c\tilde{c}(a^2+b^2+c^2+\tilde{c}^2)\nonumber \\[6pt]
&=&\varepsilon_L\varepsilon_R m_1 m_2(c^2+\tilde{c}^2)+c\tilde{c}(m_1^2+m_2^2) \nonumber \\[6pt]
&=&(cm_2+\varepsilon_L\varepsilon_R\tilde{c}m_1)(\tilde{c}m_2+\varepsilon_L\varepsilon_R cm_1)\,.
\eeqa
Hence, we end up with a remarkably simple result,
\beq \label{amusingid}
\frac{\tan\theta_L}{\tan\theta_R}=\frac{cm_2+\varepsilon_L\varepsilon_R\tilde{c}m_1}{\tilde{c}m_2+\varepsilon_L\varepsilon_R cm_1}\,.
\eeq

The case of $m_1=0$ is noteworthy.  This special case arises when $\det M =ab - c\tilde{c}=0$, in which case there is one singular
value that is equal to zero.   If $\tilde{c}\neq 0$ then inserting  $c=ab/\tilde{c}$ into \eq{Deltasqlong2} yields
$\Delta=(a^2+\tilde{c}^2)(b^2+\tilde{c}^2)/\tilde{c}^2$.  It then follows that,\footnote{One can repeat this calculation by dividing the equation $ab-c\tilde{c}=0$ by a different nonzero parameter.  For example, if $c\neq 0$ then inserting $\tilde{c}=ab/c$ into \eq{Deltasqlong2} yields
$\Delta=(a^2+c^2)(b^2+c^2)/c^2$, in which case it follows that $\tan\theta_L=c/b$ and $\tan\theta_R=a/c$, and we again recover \eq{ttspecial}.}
\beq \label{rcoslimit}
\tan\theta_L=\frac{a}{\tilde{c}}\,,\qquad\quad \tan\theta_R=\frac{\tilde{c}}{b}\,.
\eeq
In particular, after using $ab=c\tilde c$, \eq{rcoslimit} yields
\beq \label{ttspecial}
\frac{\tan\theta_L}{\tan\theta_R}=\frac{c}{\tilde{c}}\,,\qquad\text{for $m_1=0$}.
\eeq
This is indeed the correct limit of 
\eq{amusingid} when $m_1=0$, as expected.
In this case, the signs $\varepsilon_L$ and $\varepsilon_R$ are arbitrary, and  one can choose $\varepsilon_L=\varepsilon_R=1$ without loss of generality.

The case of $m\equiv m_1=m_2\neq 0$ must be treated separately.  In this case, $a=\pm b$ and $c=\mp\tilde{c}$, which yields $m=(a^2+c^2)^{1/2}$.  Since
\eq{LMR2r} implies that $MR=mL$,
one can take $R$ to be an arbitrary $2\times 2$ real orthogonal matrix.   Using \eq{eq:OLR}, the matrix $L$ is now determined,
\beq \label{realsvd}
\cos\theta_L=\frac{a\cos\theta_R-c\sin\theta_R}{\sqrt{a^2+c^2}}\,,\qquad\quad
\sin\theta_L=\pm\left(\frac{c\cos\theta_R+a\sin\theta_R}{\sqrt{a^2+c^2}}\right)\,,
\eeq
subject to $\varepsilon_L\varepsilon_R=\pm 1$, which determines the sign factor appearing in the expression for $\sin\theta_L$.

Applying the above results to $M=\left(\begin{smallmatrix} 0 & 1 \\ 1 & 0\end{smallmatrix}\right)$, we have $a=b=0$, $c=\tilde{c}=1$, $m=1$ and $\varepsilon_L\varepsilon_R=-1$.  Using \eq{realsvd}, it follows that $\cos\theta_L=-\sin\theta_R$ and $\sin\theta_L=-\cos\theta_R$.  
The corresponding singular value decomposition is given by,
\beq \label{realsvd2}
\begin{pmatrix}  -\sin\theta_R & \phm\cos\theta_R \\ \varepsilon_R\cos\theta_R & \phm\varepsilon_R\sin\theta_R\end{pmatrix}\begin{pmatrix} 0 & \phm 1 \\  1 & \phm 0\end{pmatrix}
\begin{pmatrix} \phm\cos\theta_R & \phm\varepsilon_R\sin\theta_R \\ -\sin\theta_R & \phm \varepsilon_R\cos\theta_R\end{pmatrix}=\begin{pmatrix} 1 & \phm 0 \\ 0 & \phm 1\end{pmatrix},
\eeq
which is valid for an arbitrary choice of $\theta_R$ and an arbitrary choice of sign $\varepsilon_R=-\varepsilon_L=\pm 1$.   \Eq{realsvd2} provides yet another possible form for the singular value decomposition of  $M=\left(\begin{smallmatrix} 0 & 1 \\ 1 & 0\end{smallmatrix}\right)$, to be compared with the result of \eq{LTR}.

 \section{The Autonne-Takagi factorization of a complex \texorpdfstring{$2\times 2$}{2x2} symmetric matrix}
 \label{sec:takagi}
 \bigskip

For any complex symmetric $n\times n$ matrix $M$,
there exists a unitary matrix $U$ such that,\footnote{In this section, $M$ can be either a real or complex symmetric matrix.  In the case of a real symmetric matrix~$M$, there exists a real orthogonal matrix $Q$ such that $Q^{\T}MQ={\rm diag}(m_1,m_2,\ldots m_2)$, where the $m_i$ are the eigenvalues of $M$.  The eigenvalues $m_i$ must be real, but in general they can be either positive, negative or zero.  Only in the case of a nonnegative definite real symmetric matrix $M$, where the eigenvalues $m_i$ are nonnegative, does the decomposition $Q^{\T}MQ={\rm diag}(m_1,m_2,\ldots m_2)$ constitute a Takagi diagonalization of $M$ in the space of real $n\times n$ matrices. \label{fn12}} 
\beqa
\label{app:takagi}
U^{\T} M\, U = M_D = {\rm diag}(m_1,m_2,\ldots,m_n)\,,
\eeqa
where the $m_k$ are real and non--negative.  This is the Autonne-Takagi factorization 
of the complex symmetric matrix $M$~\cite{autonne,takagi}, although this nomenclature is sometimes shortened to Takagi factorization.  
Henceforth, we shall refer to \eq{app:takagi} as the Takagi \textit{diagonalization} of a complex symmetric matrix to
contrast this with the diagonalization of
normal matrices by a unitary similarity transformation treated in Sections~\ref{sec:normal}--\ref{diagortho}.  A proof of \eq{app:takagi} is given
in Appendix D of Ref.~\cite{Dreiner:2008tw} (see also Ref.~\cite{horn}).

In general, the $m_k$ are \textit{not} the eigenvalues of $M$.
Rather, the $m_k$ are the singular values of the complex symmetric matrix
$M$. From
\eq{app:takagi} it follows that,
\beqa \label{app:diagmm}
U^\dagger  M^\dagger M U= M_D^2={\rm diag}(m^2_1,m^2_2,
     \ldots,m^2_n)\,.
\eeqa
If all of the singular values $m_k$ are non-degenerate, then one can
find a solution to \eq{app:takagi}
for $U$
from \eq{app:diagmm}.  This is no longer true if
some of the singular values are degenerate.  For example, if $M=
\bigl(\begin{smallmatrix}0\,\, & m \\ m\,\, & 0\end{smallmatrix}\bigr)$,
then the singular value $|m|$ is doubly--degenerate, but \eq{app:diagmm}
yields $U^\dagger U= \mathds{1}_{2\times 2}$, which does not
specify
$U$. That is, in the degenerate case, the Takagi diagonalization
\textit{cannot} be determined by the diagonalization of $M^\dagger M$.
Instead, one must make direct use of \eq{app:takagi}.

\Eq{app:takagi} can be rewritten as $MU=U^*M_D$, where
the columns of $U$ are orthonormal.  If we denote the $k$th
column of $U$ by $v_k$, then,
\beqa \label{mvks}
Mv_k=m_k v_k^*\,,
\eeqa
where the $m_k$ are the singular values and the vectors $v_k$
are normalized to have unit norm.   Following Ref.~\cite{takcompute},
the $v_k$ are called the {\it Takagi vectors} of the complex symmetric
$n\times n$ matrix~$M$.  

For a real symmetric matrix $M$, the Takagi
diagonalization [\eq{app:takagi}] still holds for a unitary matrix $U$,
which is easily determined as follows.  Any real symmetric
matrix $M$ can be diagonalized by a real orthogonal matrix $Z$,
\beq \label{zmz}
Z^{\T}MZ={\rm diag}(\varepsilon_1 m_1\,,\,\varepsilon_2 m_2\,,\,\ldots\,,\,
\varepsilon_n m_n)\,,
\eeq
where the $m_k$ are real and nonnegative and the $\varepsilon_k m_k$
are the real eigenvalues of $M$ with
corresponding signs
$\varepsilon_k=\pm 1$.
Then, the Takagi diagonalization of $M$ is achieved by taking
$U_{ij}=\varepsilon_i^{1/2} Z_{ij}$ (no sum over $i$).\footnote{In the case of $m_k=0$, we conventionally
choose the corresponding $\varepsilon_k=+1$.}

The Takagi diagonalization of a $2\times 2$ complex symmetric matrix can
be performed analytically.  Consider the non-diagonal complex symmetric matrix,
\beqa \label{complexsymmetric}
M=\left(\begin{array}{cc} a &\quad  c \\ c & \quad b\end{array}\right)\,,
\eeqa
where $c\neq 0$.  Following Ref.~\cite{murnaghan}, one can parameterize the 
  unitary $2\times 2$ matrix $U$ in \eq{app:takagi} as follows,
%
\beqa \label{vp}
U=VP=
\left(\begin{array}{cc} \cos\theta & \quad e^{i\phi}\sin\theta \\
-e^{-i\phi}\sin\theta &\quad \cos\theta\end{array}\right)\,
\left(\begin{array}{cc} e^{-i\alpha} & \quad 0 \\ 0 & \quad
e^{-i\beta}\end{array}\right)\,,
\eeqa
where $0\leq\theta\leq\half\pi$ and
$0\leq \alpha\,,\,\beta\,,\,\phi<2\pi$.
However, we may restrict
the angular parameter space further.  
The Takagi diagonalization equation is
\beq \label{UTMUD}
U^{\T} MU=D=\begin{pmatrix} m_1 & \,\,\, 0\\ 0 & \,\,\, m_2\end{pmatrix}\,,
\eeq
where the singular values, $m_1$  and $m_2$ are nonnegative.  
One can derive expressions for the angles $\theta$, $\phi$, $\alpha$ and $\beta$ by setting $c=\tilde{c}$, $\theta_L=\theta_R=\theta$ and $\phi_L=\phi_R=\phi$ in all results obtained in Section~\ref{sec:SVD}.   However, for pedagogical purposes, a separate derivation of the Takagi diagonalization will be presented in this section.
Using \eq{vp}, one can rewrite \eq{UTMUD} as follows,
\beq \label{VMV}
V^{\T} M V=P^*DP^*\,.
\eeq
However, $P^*DP^*$ is unchanged under the separate transformations, $\alpha\to\alpha+\pi$ and $\beta\to\beta+\pi$.
Hence, without loss of generality,
one may restrict $\alpha$ and $\beta$ to the range $0\leq\alpha\,,\,\beta<\pi$.  

Using \eq{vp}, we can rewrite 
\eq{VMV} as follows:
\beqa \label{vstar}
MV=
V^*\left(\begin{array}{cc} \sigma_1& \quad 0 \\ 0 &\quad  \sigma_2\end{array}\right)\,,
\eeqa
where
\beqa \label{msigma}
\sigma_1\equiv m_1\, e^{2i\alpha}\,,\qquad {\rm and}
\qquad  \sigma_2\equiv m_2\, e^{2i\beta}\,,
\eeqa
with real and nonnegative $m_1$ and $m_2$.   The singular values of $M$ can be derived by taking the nonnegative square roots of the eigenvalues of $M^\dagger M$,
\beq  \label{mk2}
m_{1,2}^2=|\sigma_{1,2}|^2=\tfrac{1}{2}\left[|a|^2+|b|^2+2|c|^2\mp
\widetilde{\Delta}\right]\!,
\eeq
in a convention where $0\leq m_1\leq m_2$ (i.e., $\widetilde{\Delta}\geq 0$), with
\beqa
\widetilde\Delta &\equiv &\bigl[(|a|^2-|b|^2)^2
                    +4|a^*c+b c^*|^2\bigr]^{1/2}\nonumber \\[6pt]
                    &=& \bigl[(|a|^2+|b|^2+2|c|^2)^2
                    -4|a b - c^2|^2\bigr]^{1/2}\,. \label{DeltaTakagi}
\eeqa

To evaluate the angles $\phi$ and $\theta$ (which determine the matrix $V$), we multiply
out the matrices in \eq{vstar}.  The end result is,
\beqa
\sigma_1 &=& a- c\, e^{-i\phi}\tan\theta
= b\, e^{-2i\phi}-c\, e^{-i\phi}\cot\theta\,,\label{sig1}\\[6pt]
\sigma_2 &=& b+ c\, e^{i\phi}\tan\theta \,\,\,\,=
a\, e^{2i\phi}+c\, e^{i\phi}\cot\theta\,.\label{sig2}
\eeqa

We first assume that $m_1\neq m_2$, corresponding to the case of nondegenerate singular values of $M$.  
Using either \eq{sig1} or (\ref{sig2}), and making use of the trigonometric identity, 
\beq \label{trigid}
\tan 2\theta=2(\cot\theta-\tan\theta)^{-1}\,,
\eeq 
one obtains a simple equation for
$\tan 2\theta$,
\beqa \label{t2th}
\tan 2\theta=\frac{2c}{b\, e^{-i\phi}-a\,e^{i\phi}}\,.
\eeqa
Since $\tan 2\theta$ is real, it follows that
\beq \label{imbc}
\Im(bc^*\, e^{-i\phi}-ac^*\, e^{i\phi})=0\,.
\eeq
One can then use \eq{imbc} to obtain an expression for $e^{2i\phi}$,
\beqa \label{eiphi2}
e^{2i\phi}=\frac{a^*c+bc^*}{ac^*+b^*c}\,,
\eeqa
or equivalently,
\beqa\label{eiphi}
e^{i\phi}=\frac{\varepsilon(a^*c+bc^*)}{|a^*c+bc^*|}\,,\quad \text{where $\varepsilon=\pm 1$}.
\eeqa
The choice of sign in \eq{eiphi} is determined by our convention that $m_1< m_2$ (in the nondegenerate case) or equivalently, $|\sigma_1|^2< |\sigma_2|^2$.
Thus, to determine $\varepsilon$, we make use of \eqs{sig1}{sig2} to obtain two different expressions for $|\sigma_2|^2-|\sigma_1|^2$,
\beqa
|\sigma_2|^2-|\sigma_1|^2 &=& |b|^2-|a|^2+\bigl[(ac^*+b^* c)e^{i\phi}+(a^* c+bc^*)e^{-i\phi}\bigr]\tan\theta  \nonumber \\[6pt]
&=& |a|^2-|b|^2+\bigl[(ac^*+b^* c)e^{i\phi}+(a^* c+bc^*)e^{-i\phi}\bigr]\cot\theta\,.  \label{sig21}
\eeqa
Using \eq{eiphi} to eliminate $\phi$, it follows that
\beq \label{sig21two}
|\sigma_2|^2-|\sigma_1|^2 = |b|^2-|a|^2+2\varepsilon|a^*c+bc^*|\tan\theta= |a|^2-|b|^2+2\varepsilon|a^*c+bc^*|\cot\theta\,.
\eeq
Adding the two expressions given in \eq{sig21two} for $|\sigma_2|^2-|\sigma_1|^2$, we end up with
\beq \label{sigdiff}
|\sigma_2|^2-|\sigma_1|^2=\varepsilon|a^*c+bc^*|(\tan\theta+\cot\theta)\,.
\eeq
Since $|\sigma_2|^2> |\sigma_1|^2$ and $0\leq\theta\leq\half\pi$, it follows that $\varepsilon=1$.  Moreover, \eq{sigdiff} implies that in the case of nondegenerate singular values, $a^*c+bc^*\neq 0$.  This latter condition ensures that none of the denominators in \eqss{t2th}{eiphi2}{eiphi} vanish.

We can now obtain an explicit form for $\tan 2\theta$ by either subtracting the two expressions given in \eq{sig21two} for $|\sigma_2|^2-|\sigma_1|^2$ or by
inserting the result for $e^{i\phi}$ back into \eq{t2th}.  Taking into account that $\varepsilon=1$, both methods yield the same final result,
\beqa \label{tantwotheta}
\tan 2\theta=\frac{2|a^*c+bc^*|}{|b|^2-|a|^2}\,.
\eeqa
Using \eqs{trigid}{tantwotheta}, it follows that
\beq \label{tanth3} 
\tan\theta=\frac{|a|^2-|b|^2+\widetilde{\Delta}}{2|a^*c+bc^*|}\,,\qquad\quad \cot\theta = \frac{|b|^2-|a|^2+\widetilde{\Delta}}{2|a^*c+bc^*|}\,.
\eeq
If we now insert the results of \eq{tanth3}  into \eq{sigdiff} with $\varepsilon=1$, it then follows that,
\beq \label{msqdiff}
|\sigma_2|^2-|\sigma_1|^2 =\widetilde{\Delta}\,.
\eeq
One can quickly compute $|\sigma_1|^2+|\sigma_2|^2$ by noting that,
\beq \label{msqsum}
|\sigma_1|^2+|\sigma_2|^2=m_1^2+m_2^2={\rm Tr}(M^\dagger M)=|a|^2+|b|^2+2|c|^2\,.
\eeq
Adding and subtracting \eqs{msqdiff}{msqsum} reproduces the expressions of $m_{1,2}^2=|\sigma_{1,2}|^2$ obtained in \eq{mk2}.

It is sometimes more convenient to rewrite \eq{tanth3} in another form,
\beq
\tan^2\theta=\frac{\widetilde{\Delta}+|a|^2-|b|^2}{\widetilde{\Delta}-|a|^2+|b|^2}\,.
\eeq
If we now make use of the trigonometric identity, $\cos 2\theta=(1-\tan^2\theta)/(1+\tan^2\theta)$, we end up with a rather simple expression,
\beq
\cos 2\theta=\frac{|b|^2-|a|^2}{\widetilde{\Delta}}\,.
\eeq
One can now use this result to derive,
\beq
\cos\theta=\sqrt{\frac{\widetilde\Delta-|a|^2+|b|^2}{2\widetilde\Delta}}\,,\qquad\quad  \sin\theta= \sqrt{\frac{\widetilde\Delta+|a|^2-|b|^2}{2\widetilde\Delta}}\,.
\eeq

The final step of the computation is the determination of the angles
$\alpha$ and $\beta$ from \eq{msigma}.  Employing \eq{tanth3} together with \eq{eiphi} with $\varepsilon=1$ and \eq{mk2}, one can establish the following useful results,
\beq \label{useful}
e^{-i\phi}\tan\theta=\frac{a c^*+b^*c}{|b|^2+|c|^2-|\sigma_1|^2}\,,\qquad\quad
e^{i\phi}\tan\theta=\frac{a^* c+bc^*}{|\sigma_2|^2-|a|^2-|c|^2}\,.
\eeq
Inserting \eq{useful} into
\eqs{sig1}{sig2} yields,
\beqa
\sigma_1 &=& m_1 e^{2i\alpha}= a- c\, e^{-i\phi}\tan\theta =\frac{a\bigl(|b|^2-|\sigma_1|^2\bigr)-b^* c^2}{|b|^2+|c|^2-|\sigma_1|^2}\,, \\[6pt]
\sigma_2 &=& m_2 e^{2i\beta}= b+ c\, e^{i\phi}\tan\theta =\frac{b\bigl(|\sigma_2|^2-|a|^2\bigr)+a^* c^2}{|\sigma_2|^2-|a|^2-|c|^2}\,.
\eeqa
Hence, it immediately follows that,
\beqa
\alpha &=& \half\arg\bigl\{a\bigl(|b|^2-m_1^2\bigr)-b^*c^2\bigr\}
\label{alphadef}\,,\\[6pt]
\beta &=& \half\arg\bigl\{b\bigl(m_2^2-|a|^2\bigr)+a^*c^2\bigr\}\,.
\label{betadef}
\eeqa

The case of $m_1=0$ is noteworthy.  This special case arises when 
${\rm det}~M=ab-c^2=0$, in which case there is one singular
value that is equal to zero.  In particular, it then follows that $\widetilde{\Delta}=(|a|+|b|)^2$ [cf.~\eq{DeltaTakagi}] and
$m_2^2={\rm Tr}(M^\dagger M)=|a|^2+|b|^2+2|c|^2$.  Inserting $c^2=ab$ in the latter expression yields
$m_2=|a|+|b|$.   In addition, 
\beq
\tan\theta=\left|a/b\right|^{1/2}\,,\qquad\quad \phi=\arg(b/c)=\arg(c/a)\,,\qquad\quad \beta=\half\arg b\,.
\eeq
However, $\alpha$ is undefined, since the argument of
\eq{alphadef} vanishes.  This corresponds to
the fact that for a zero singular value, the corresponding
(normalized) Takagi vector is only unique up to an overall arbitrary
phase.\footnote{The normalized Takagi vectors are
unique up to an overall sign if the corresponding singular values are
non--degenerate and non--zero.  However, in the case of
a zero singular value or a pair of degenerate of
singular values, there is more freedom in defining the Takagi
vectors.  For further details, see Appendix D of Ref.~\cite{Dreiner:2008tw}. \label{fn}}
One can now check that all the results obtained above agree with the corresponding results of Section~\ref{sec:SVD} after making the substitutions, $\tilde{c}=c$, $\theta_{L,R}=\theta$ and $\phi_{L,R}=\phi$, as previously noted.

We provide one illuminating example of the above results.  Consider
the complex symmetric matrix,
\beqa \label{mex}
M=\left(\begin{array}{cc} 1 &  \,\,\,\phm i\\ i & \,\,\, -1
\end{array}\right)\,.
\eeqa
The eigenvalues of $M$ are degenerate and equal to zero.  However,
there is only one linearly independent eigenvector, which is proportional
to $(1\,,\,i)$.  Thus, $M$ cannot be diagonalized by a similarity
transformation.
In contrast, all complex symmetric matrices are Takagi-diagonalizable.
The singular values of $M$ are 0 and 2 (since these
are the non--negative square roots of the eigenvalues of $M^\dagger M$),
which are \textit{not} degenerate.  Thus, all the formulae derived above
apply in this case.  One quickly determines that $\theta=\tfrac14\pi$,
$\phi=\half\pi$, $\beta=\half\pi$ and $\alpha$ is indeterminate.  The resulting Takagi diagonalization is
$U^{\T}MU={\rm diag}(0\,,\,2)$ with:
\beqa
U=\frac{1}{\sqrt{2}}\left(\begin{array}{cc} 1 & \,\,\,\phm i\\
      i &  \,\,\,\phm 1 \end{array}\right)\,\left(\begin{array}{cc} e^{-i\alpha} &\,\,\,
      \phm 0\\ 0 & \,\,\, -i \end{array}\right)
= \frac{1}{\sqrt{2}}\left(\begin{array}{cc} e^{-i\alpha} &\,\,\, \phm 1\\
     i e^{-i\alpha} &\,\,\,  -i \end{array}\right)\,.
\eeqa
Thus, $U$ is unique up to an overall factor of $-1$ and an arbitrary phase $\alpha$.  The latter is a consequence of the presence of
a zero singular value.
This example illustrates the distinction between the
(absolute values of the) eigenvalues of $M$ and its singular values.
It also exhibits the fact that one cannot always perform a Takagi
diagonalization by computing the eigenvalues and eigenvectors of $M^\dagger M$.

Finally, we treat the case of degenerate nonzero singular
values, i.e.~$m\equiv m_1=m_2\neq 0$.  
As indicated below \eq{complexsymmetric}, we shall continue to 
assume that $c\neq 0$.  In light of \eq{sigdiff}, the degenerate case arises when
\beq \label{app:bcac}
a^*c+bc^*=0\,.
\eeq
If \eq{app:bcac} is satisfied, then it follows from \eq{mk2} that
\beq
m=m_1=m_2=\sqrt{|b|^2+|c|^2}\,.
\eeq
Moreover, 
$\phi$ and $\theta$ are indeterminate in light of \eqs{eiphi2}{tantwotheta}.   Nevertheless, these two indeterminate angles are related if $a$, $b\neq 0$.  Using \eqss{sig1}{sig2}{app:bcac}, it follows that,
\beq \label{singtan}
 \tan 2\theta=\bigl[\Re(b/c)c_\phi+\Im(b/c)s_\phi\bigr]^{-1}\,, 
\eeq
where $c_\phi\equiv\cos\phi$ and $s_\phi\equiv\sin\phi$.   In contrast to \eq{imbc}, the reality of $\tan 2\theta$ imposes no constraint on $\phi$  
in the case of degenerate singular values.  Consequently, the angle $\phi$ is indeed indeterminate.\footnote{The same conclusion also follows from \eq{UTMUD}.
If $D=m\mathds{1}_{2\times 2}$ then $(U{\cal O})^{\T}M(U{\cal O})
={\cal O}^{\T}D{\cal O}=D$ for any real orthogonal matrix ${\cal O}$.
In particular, $\phi$ simply represents the freedom to choose
${\cal O}$ [cf.~\eq{uspecial}].}  
Since $\phi$ is indeterminate, \eq{singtan} implies that $\theta$ is
indeterminate as well, except in the special case of $a=b=0$.   In this latter case, \eq{app:bcac} is satisfied and the singular values of $M$ are degenerate.   However, \eq{singtan} 
does not relate $\theta$ to the indeterminate angle~$\phi$.  Indeed, \eq{sig1} yields $\theta=\tfrac14\pi$, which is also consistent with the $b\to 0$ limit of \eq{singtan}. 

In the case of degenerate singular values,
 \eqs{alphadef}{betadef} are no longer valid, as their derivation relies on the results of \eqs{eiphi}{tanth3}, which are indeterminate expressions when $a^*c+bc^*=0$.   Hence, we need another technique to determine the angles $\alpha$ and $\beta$.   Employing \eqss{sig1}{sig2}{app:bcac} we can derive the following results after some manipulations,
 \beqa
\sigma_1&=& m e^{2i\alpha} = -c\,e^{-i\phi}\bigl[(1+A^2)^{1/2}+iB\bigr]\, \label{sig1p} \\[3pt]
\sigma_2&=& m e^{2i\beta} = c\,e^{i\phi}\bigl[(1+A^2)^{1/2}-iB\bigr]\,,\label{sig2p}
\eeqa
where $m=(|b|^2+|c|^2)^{1/2}$ and
\beq
A\equiv \Re(b/c)c_\phi+\Im(b/c)s_\phi\,,\qquad B\equiv \Re(b/c)s_\phi-\Im(b/c)c_\phi\,.
\eeq
Thus, the angles $\alpha$ and $\beta$ are separately determined by \eqs{sig1p}{sig2p} in terms of the indeterminate angle $\phi$.   Nevertheless, the sum $\alpha+\beta$ is independent of~$\phi$.   This is most easily seen by employing \eqs{sig1p}{sig2p} to obtain,
\beq
c\sigma_1^*+c^*\sigma_2=0\,.
\eeq
Hence, it follows that,
\beq \label{app:ccstar}
e^{2i(\alpha+\beta)}=-\frac{c}{c^*}\,.
\eeq
Thus, the matrix $U$ in \eq{UTMUD} is now fixed in terms of the quantity $\alpha+\beta$ and the indeterminate angle $\phi$.  

We illustrate the above results with the example of
$M=\bigl(\begin{smallmatrix}0 & 1 \\ 1 & 0\end{smallmatrix}\bigr)$.\footnote{This example is of particular interest to physicists, since the matrix $mM$ (for positive number $m$)
corresponds to the mass matrix of a Dirac fermion of mass $m$ that arises when expressed in a basis of two-component spinors.   The Takagi diagonalization of $mM$ demonstrates that a Dirac fermion of mass $m$  is physically equivalent to two mass-degenerate Majorana fermions of mass $m$.  Further details can be found in Ref.~\cite{Dreiner:2008tw}.}
In this case
$M^\dagger M=\mathds{1}_{2\times 2}$, so $U$ cannot be deduced by
diagonalizing $M^\dagger M$.  Setting $a=b=0$ and $c=1$ in the above
formulae, it follows that $m=1$, $\theta=\tfrac14\pi$,
$\sigma_1=-e^{-i\phi}$ and $\sigma_2=e^{i\phi}$, which yields
$\alpha=-\half(\phi\pm\pi)$ and $\beta=\half\phi$.  Thus, \eq{vp} yields,
\beqa \label{uspecial}
U&=&\frac{1}{\sqrt{2}}\left(\begin{array}{cc} \phm 1 & \quad  e^{i\phi} \\
-e^{-i\phi} & \quad 1
\end{array}\right)\,  \left(\begin{array}{cc}\pm  ie^{i\phi/2} & \quad 0  \\
  0 & \quad  e^{-i\phi/2} \end{array}\right)  =
\frac{1}{\sqrt{2}}\left(\begin{array}{cc} \pm ie^{i\phi/2} & \quad 
e^{i\phi/2} \\ \mp i e^{-i\phi/2} & \quad e^{-i\phi/2}
\end{array}\right)\nonumber \\[10pt]
&=&\frac{1}{\sqrt{2}}\left(\begin{array}{cc} \phm i& \quad 1 \\ -i & \quad 1
\end{array}\right)\,\left(\begin{array}{cc}
\pm \cos(\phi/2) & \quad \sin(\phi/2) \\
 \mp\sin(\phi/2) & \quad \cos(\phi/2)\end{array}\right)\,,
\eeqa
which shows that in the case of degenerate singular values, $U$ is unique only up to multiplication on
the right by an arbitrary orthogonal matrix. 

For completeness, it is instructive to examine the special case of the Takagi diagonalization of a non-diagonal \textit{real} symmetric matrix $M=\left(\begin{smallmatrix} a & c \\ c & b\end{smallmatrix}\right)$, where $c\neq 0$. In this case, the singular values, $m_1$ and $m_2$ are the nonnegative square roots of
\beq  \label{mk2r}
m_{1,2}^2=\tfrac{1}{2}\left[a^2+b^2+2c^2\mp
\widetilde{\Delta}\right]\!,
\eeq
where
\beq
\widetilde\Delta \equiv |a+b|\bigl[(a-b)^2
                    +4c^2\bigr]^{1/2}
                    = \bigl[(a^2+b^2+2c^2)^2
                    -4(a b - c^2)^2\bigr]^{1/2}\,. \label{RealDeltaTakagi}
\eeq
in a convention where $0\leq m_1\leq m_2$.  Assuming that $m_1\neq m_2$, the latter implies that one must take $\varepsilon=1$ in \eq{eiphi}, which yields
\beq
\phi=\begin{cases} 0\,, & \text{if $\sgn\bigl(c(a+b)\bigr)=+1$}\,, \\
\pi\,, &  \text{if $\sgn\bigl(c(a+b)\bigr)=-1$}\,.\end{cases}
\eeq
It is therefore convenient to redefine $\theta\to \theta\sgn\bigl(c(a+b)\bigr)$, in which case $-\half\pi<\theta\leq\half\pi$.  
Then, the Takagi diagonalization of~$M$ is given by \eq{UTMUD}, where
\beqa \label{vpreal}
U=
\left(\begin{array}{cc} \phm\cos\theta & \quad \sin\theta \\
-\sin\theta &\quad \cos\theta\end{array}\right)\,
\left(\begin{array}{cc} e^{-i\alpha} & \quad 0 \\ 0 & \quad
e^{-i\beta}\end{array}\right)\,,
\eeqa
and the redefined angle $\theta$ is given by,
\beq \label{tanth3real} 
\tan\theta=\frac{\widetilde{\Delta}+a^2-b^2}{2c(a+b)}\,.
\eeq
It then follows that
\beq
\cos\theta=\sqrt{\frac{\widetilde\Delta-a^2+b^2}{2\widetilde\Delta}}\,,\qquad\quad  \sin\theta= \sgn\bigl(c(a+b)\bigr)\sqrt{\frac{\widetilde\Delta+a^2-b^2}{2\widetilde\Delta}}\,.
\eeq
Finally, one can obtain compact expressions for the angles $\alpha$ and $\beta$ using \eqs{alphadef}{betadef},
\beq
 \alpha=\begin{cases} \,\,0\,, & \text{if $\sgn\bigl(b\det M-am_1^2\bigr)=+1$}, \\
 \half\pi\,, &  \text{if $\sgn\bigl(b\det M-am_1^2\bigr)=-1$}, \end{cases}
 \qquad\quad
 \beta=\begin{cases} \,\,0\,, & \text{if $\sgn\bigl(bm_2^2-a\det M\bigr)=+1$}, \\
 \half\pi\,, &  \text{if $\sgn\bigl(bm_2^2-a\det M\bigr)=-1$}. \end{cases}
\eeq
In the special case of $m_1=0$, we have $ab=c^2\neq 0$, in which case the angle $\alpha$ is indeterminate and
$\beta=0$ [$\half\pi$] for $b>0$ [$b<0$].  Henceforth, we shall assume that $m_1>0$.

Considering that $\det M=ab-c^2=\xi m_1 m_2$, where $\xi\equiv\sgn(ab-c^2)$, it then follows that
\beq \label{inequalities}
 \alpha=\begin{cases} \,\,0\,, & \text{if $\sgn\bigl(\xi bm_2-am_1\bigr)=+1$}, \\
 \half\pi\,, &  \text{if $\sgn\bigl(\xi bm_2-am_1\bigr)=-1$}, \end{cases}
 \qquad\quad
 \beta=\begin{cases} \,\,0\,, & \text{if $\sgn\bigl(\xi bm_2- am_1\bigr)=+\xi$}, \\
 \half\pi\,, &  \text{if $\sgn\bigl(\xi bm_2- am_1\bigr)=-\xi$}. \end{cases}
\eeq
That is, the matrix $U$ is real and orthogonal (corresponding to $\alpha=\beta=0$) if and only if $ab\geq c^2$
and $bm_2>am_1$.  In Appendix B, we show that $ab\geq c^2$ and $bm_2>am_1$ are both satisfied if and only if $\det M\geq 0$ and $\Tr M>0$.   In particular, we can identify $m_1$ and $m_2$ as the two eigenvalues of $M$.
Hence, in this case the diagonalization of $M$ by a real orthogonal matrix given in Section~\ref{diagortho} constitutes a Takagi diagonalization of 
$M$ [cf.~footnote~\ref{fn12}].  

In the case of $m_1=m_2$, it follows that $a=-b$, 
so that $\det M<0$.  Indeed,
\eq{app:ccstar} yields $\alpha+\beta=\half\pi$, which implies that the Takagi diagonalization 
matrix $U$ is not real, as expected.

\section*{Acknowledgments}

Some aspects of the computations of the singular value decomposition of a complex  $2\times 2$ matrix and the
Autonne-Takagi factorization of a complex $2\times 2$ symmetric matrix were carried out
in collaboration with Seong Youl Choi.  I also gratefully acknowledge my co-authors, Herbi Dreiner and Stephen 
Martin, of the review article cited in Ref.~\cite{Dreiner:2008tw}.   Some of the material of these notes has been taken from Appendix D of Ref.~\cite{Dreiner:2008tw}.
H.E.H. is supported in part by the U.S. Department of Energy Grant
No.~\uppercase{DE-SC}0010107.

\begin{appendices}

\section{Singular value decomposition of a matrix with degenerate singular values revisited}
\renewcommand{\theequation}{A.\arabic{equation}}
\setcounter{equation}{0}

Recall that the singular value decomposition of the $2\times 2$ matrix $M=\left(\begin{smallmatrix} a & c \\ \tilde{c} & b\end{smallmatrix}\right)$ with two degenerate singular values given by $m=\sqrt{|a|^2+|c|^2}$ is,
\beq \label{app:LMR22}
L^{\T} MR=m\mathds{1}_{2\times 2}\,. 
\eeq
In general we can parameterize
two $2\times 2$ unitary matrices $L$ and $R$ in \eq{app:svd} by %
\beqa
&& L = U_L P_L
     = \left(\begin{array}{cc}
         \cos\theta_L   &    e^{i \phi_L} \sin\theta_L  \\
     -e^{-i\phi_L} \sin\theta_L & \cos\theta_L
             \end{array}\right) \,
       \left(\begin{array}{cc}
          e^{-i\alpha_L}  &  0 \\
          0  & e^{-i\beta_L}
             \end{array}\right)\,,\label{app:ulpl} \\[10pt]
&& R = U_R P_R
     = \left(\begin{array}{cc}
         \cos\theta_R   &    e^{i \phi_R} \sin\theta_R  \\
     -e^{-i\phi_R} \sin\theta_R & \cos\theta_R
       \end{array}\right)\,
       \left(\begin{array}{cc}
          e^{-i\alpha_R}  &  0 \\
          0  & e^{-i\beta_R}
             \end{array}\right)\,,\label{app:urpr}
\eeqa
Here, we will allow the phase matrices $P_L$ and $P_R$ to be different, although in the end only $\alpha_L+\alpha_R$ and $\beta_L+\beta_R$ are fixed by \eq{app:LMR22}.

Consider the case of degenerate singular values treated in Section~\ref{sec:SVD}.   If $P_L\neq P_R$, then \eqst{CL2}{cctil} are slightly modified,
\beqa
\hspace{-0.5in}
m\cos\theta_L &=&  e^{-i(\alpha_L+\alpha_R)}\bigl(a\cos\theta_R-c e^{-i\phi_R}\sin\theta_R\bigr) = -\frac{\tilde{c}^*}{c}e^{i(\beta_L+\beta_R)}\bigl(a\cos\theta_R-ce^{-i\phi_R}\sin\theta_R\bigr),  \label{app:CL2} \\[6pt]  \hspace{-0.5in}
m e^{i\phi_L}\sin\theta_L &=& \frac{\tilde{c}^*}{c}e^{i(\beta_L+\beta_R)}\bigl(\tilde{c}\cos\theta_R-be^{-i\phi_R}\sin\theta_R\bigr)=-e^{-i(\alpha_L+\alpha_R)}\bigl(\tilde{c}\cos\theta_R-be^{-i\phi_R}
\sin\theta_R\bigr).\label{app:SL2}
\eeqa
Since both \eqs{CL2}{SL2} cannot simultaneously vanish, it follows that 
\beq \label{app:cctil}
e^{i(\alpha_L+\alpha_R+\beta_L+\beta_R)}=-\frac{c}{\tilde{c}^*}\,.
\eeq

 As previously noted in \eq{degenconds},
degenerate singular values exist if and only if 
\beq \label{app: degenconds}
|a|=|b|\,,\, |c|=|\tilde{c}|\,,\, \text{and $a^* c = - b \tilde{c}^*$}.
\eeq  
\Eq{app: degenconds} also implies that $a^*\tilde{c}=-bc^*$.
By re-expressing $b$ in terms of $a, c$ and $\tilde{c}$,
one can cast the matrix $M$ in the form,
\beqa
M  &=& \left(\begin{array}{cc}
                |a|\, e^{i\phi_a}  & \quad |c|\, e^{i\phi_c}  \\
                |c|\,\, e^{i\phi_{\tilde{c}}}
                &\quad  -|a|\, e^{i(\phi_c+\phi_{\tilde{c}}-\phi_a)}
                \end{array}\right)
   =  \left(\begin{array}{cc}
            e^{i\phi_a/2}   & \,\,\,  0   \\
            0  & \,\,\,e^{i(\phi_{\tilde{c}}-\phi_a/2)}
            \end{array}\right)
      \left(\begin{array}{cr}
               |a|   & \,\,\,\phm |c|  \\
               |c|   & \,\,\,-|a|
            \end{array}\right)
      \left(\begin{array}{cc}
            e^{i\phi_a/2}   & \,\,\,0   \\
            0  & \,\,\, e^{i(\phi_c-\phi_a/2)}
            \end{array}\right), \nonumber \\
 &&\phantom{line}           
\label{eq:degenerate_matrix_simplified}
\eeqa
where $a\equiv|a|e^{i\phi_a}$, $c\equiv|c|e^{i\phi_c}$ and $\tilde{c}\equiv|c|e^{i\phi_{\tilde{c}}}$ (after making use of $|c|=|\tilde{c}|$).

One possible choice for the singular value decomposition of $M$ [\eq{app:LMR22}] is to employ the unitary matrices
\beq \label{app:LR}
L=  
\begin{pmatrix}  e^{-i\phi_a/2} & \,\,\, 0 \\  0& \,\,\,
e^{-i(\phi_{\tilde{c}}-\phi_a/2)} \end{pmatrix} \, QP\,,\qquad\quad R= \begin{pmatrix}  e^{-i\phi_a/2} & \,\,\, 0 \\  0 & \,\,\, e^{-i(\phi_c-\phi_a/2)}\end{pmatrix}\, QP\,,
\eeq
where $Q$ is a real orthogonal matrix and $P$ is a $2\times 2$ diagonal phase matrix $P={\rm diag}(i\,,\,1)$.  Then, \eq{LMR22}
yields
\beq
Q^{\T}  \left(\begin{array}{cr}
               |a|   & \,\phm |c|  \\
               |c|   & \,-|a|
            \end{array}\right) Q = P^* \begin{pmatrix} m & \,\,\,  0 \\  0 & \,\,\, m\end{pmatrix}P^*= \begin{pmatrix} -m & \,\,\,  0 \\  \phm 0 & \,\,\, m\end{pmatrix}\,,
 \eeq       
 where
\beq \label{app:msing}
m=\sqrt{|a|^2+|c|^2}\,.
\eeq
That is, $Q$ is the real orthogonal matrix that diagonalizes the real symmetric matrix, $\left(\begin{smallmatrix} |a| & \phm |c| \\ |c| & -|a| \end{smallmatrix}\right)$,
whose eigenvalues are $\lambda_{1,2}=-m$, $m$ (whereas its singular values are degenerate and equal to~$m$).
The explicit form for $Q$ can be determined using the results of Section~\ref{diagortho}.

Hence, one possible choice for the singular value decomposition of $M$ takes the following form in the case degenerate singular values,
\beqa
m\id &=& L^{\T}MR=P^{\T} Q^{\T} \left(\begin{array}{cr}
            |a|   &\,\,\,\phm  |c|  \\
            |c|   &\,\,\, -|a|
              \end{array}\right) Q P \nonumber\\[8pt]
 &=&
\left(\begin{array}{cc}
          i  & \quad  0  \\
          0   &   \quad 1
            \end{array}\right)
          \left(\begin{array}{cc}
  \cos\theta  &  \,\,\, -\sin\theta  \\
  \sin\theta  &   \,\,\, \phm \cos\theta
       \end{array}\right)  
\left(\begin{array}{cr}
            |a|   & \,\,\, \phm  |c|  \\
            |c|   & \,\,\,  -|a|
              \end{array}\right)
\left(\begin{array}{cc}
       \phm \cos\theta  &    \,\,\, \sin\theta  \\
      - \sin\theta  &    \,\,\, \cos\theta
       \end{array}\right)
\left(\begin{array}{cc}
          i  &     \quad 0  \\
          0   &      \quad 1
      \end{array}\right)\,,
\label{eq:degenerate_svd_diag}
\eeqa
where
the rotation angle $\theta$ of the orthogonal matrix $Q$ is given by [cf.~\eqst{Udiag}{quad4}],
\beq \label{degentheta}
\cos\theta=\sqrt{\frac{1-|a|/m}{2}}\,,\qquad\qquad \sin\theta=\sqrt{\frac{1+|a|/m}{2}}\,.
\eeq

It is instructive to check that \eqs{eq:degenerate_svd_diag}{degentheta} are consistent with the general form of the singular value decomposition in the degenerate case obtained in 
\eqst{app:CL2}{app:cctil}.   If we compare \eq{app:LR} with the forms for $L$ and $R$ given in \eqs{app:ulpl}{app:urpr}, we can identify,
\beqa
&& \theta_L=\theta_R\,,\qquad   \alpha_L=\alpha_R=\half(\phi_a-\pi)\,,\qquad \beta_L=\phi_{\tilde{c}}-\half\phi_a\,,\qquad \beta_R=\phi_c-\half\phi_a\,,\nonumber \\
&& \phi_L=\phi_{\tilde{c}}-\phi_a\,,\qquad \phi_R=\phi_c-\phi_a\,. \label{app:special}
\eeqa
Note that by inserting $c=|c|e^{i\phi_c}$ and $\tilde{c}=|c|e^{i\phi_{\tilde{c}}}$ into \eq{app:cctil}, it follows that
\beq
\alpha_L+\alpha_R+\beta_L+\beta_R=\phi_c+\phi_{\tilde{c}}-\pi\,,
\eeq
which is consistent with \eq{app:special}.

Finally, we insert \eq{app:special} into \eqs{app:CL2}{app:SL2} to obtain,
\beqa
m\cos\theta&=& |c|\sin\theta-|a|\cos\theta\,, \\
m\sin\theta&=& |a|\sin\theta+|c|\cos\theta\,,
\eeqa
where $\theta\equiv\theta_L=\theta_R$.  Both equations above are consistent, in light of \eq{app:msing}, and yield
\beq
\tan\theta =\frac{|c|}{m-|a|}=\frac{\sqrt{m^2-|a|^2}}{m-|a|}=\sqrt{\frac{m+|a|}{m-|a|}}\,,
\eeq
which coincides with the result of \eq{degentheta}.

Of course, \eq{eq:degenerate_svd_diag} is not the most general singular value decomposition of $M$ in the case of degenerate singular values, since we are free to choose a more general form for $R$ that would yield $\theta_L\neq \theta_R$.  For example, it is possible to choose $L=\mathds{1}_{2\times 2}$.  To see that this is a consistent choice, we plug this result back into 
\eq{app:LMR22} to obtain 
\beq \label{emare}
MR=m\id\,.
\eeq
Multiplying this equation by its adjoint yields, 
\beq \label{ememdag1}
MM^\dagger=M^\dagger M=m^2\id\,.
\eeq
By explicit computation with $M=\left(\begin{smallmatrix} a & c \\ \tilde{c} & b\end{smallmatrix}\right)$,
\beq \label{ememdag2}
MM^\dagger=M^\dagger M=(|a|^2+|c|^2)\id\,,
\eeq
after making use of \eq{app: degenconds}.  Indeed, \eqs{ememdag1}{ememdag2} are equivalent in light of \eq{app:msing}.   Therefore, it follows that $M^\dagger=m^2 M^{-1}$.
Inserting this last result into \eq{emare}, we conclude that one of the singular value decompositions of $M$ in the case of degenerate singular values is given by
\beq
L^{\T} MR=m\mathds{1}_{2\times 2}\,,\qquad\quad \text{where $L=\id$ and $R=\frac{1}{m}M^\dagger$}.
\eeq
By a similar argument, one can obtain another singular value decompositions of $M$ in the case of degenerate singular values by taking $R=\id$, which yields
 \beq
L^{\T} MR=m\mathds{1}_{2\times 2}\,,\qquad\quad \text{where $L=\frac{1}{m}M^*$ and $R=\id$}.
\eeq

\section{On the Takagi diagonalization of a real \texorpdfstring{$2\times 2$}{2x2} symmetric matrix}
\renewcommand{\theequation}{B.\arabic{equation}}
\setcounter{equation}{0}

At the end of Section~\ref{sec:takagi}, we considered the Takagi diagonalization of a real symmetric matrix, $U^T M U={\rm diag}(m_1,m_2)$, where $m_1$ and $m_2$ are the singular values of $M$ (which are nonnegative quantities).  Thus the Takagi diagonalization of $M=\left(\begin{smallmatrix} a & c \\ c & b\end{smallmatrix}\right)$ differs from the diagonalization of $M$ treated in Section~\ref{diagortho} unless the eigenvalues of $M$ are nonnegative.  One consequence of \eq{inequalities} is that the Takagi diagonalization matrix $U$ is a real orthogonal matrix if and only if $ab\geq c^2\neq 0$ and $bm_2>am_1$.   In this Appendix, we shall verify this last assertion.

Since $ab\geq c^2\neq 0$, then $a$ and $b$ are either both positive or both negative.   First, assume that $a$, $b>0$.   Then, the condition $bm_2>am_1$ is equivalent to the condition that $(m_2/m_1)^2>(a/b)^2$.   Employing \eq{mk2r}, it follows that
\beq
b^2\bigl[a^2+b^2+2c^2+\widetilde{\Delta}\bigr]>a^2\bigl[a^2+b^2+2c^2-\widetilde{\Delta}\bigr]\,,
\eeq
which yields
\beq
(a^2+b^2)\widetilde{\Delta}>(a^2-b^2)(a^2+b^2+2c^2)\,.
\eeq
This equality is trivially satisfied if $a\leq b$, so let us assume that $a>b$.  Then, one can square both sides of the inequality above to obtain,
\beq
(a^2+b^2)^2\bigl[(a^2+b^2+2c^2)^2-4(ab-c^2)^2\bigr]-(a^2-b^2)^2(a^2+b^2+2c^2)^2>0\,.
\eeq
After some algebraic manipulations, the end result is
\beq
4c^2(a+b)^2\bigl[ab(a+b)^2+(ab-c^2)(a-b)^2\bigr]>0\,,
\eeq
which is manifestly true given that $a$, $b>0$ and $ab\geq c^2$.

Second, assume that $a$, $b<0$.  Then, the condition $bm_2>am_1$ is equivalent to the condition that $(m_2/m_1)^2<(a/b)^2$.  Following the same steps as above, one obtains inequalities that are never satisfied.   Hence, one can conclude that if $ab\geq c^2$, then $bm_2>am_1$ is satisfied if and only if $a$, $b>0$.   Finally, the conditions $ab\geq c^2$ and $a$, $b>0$ are equivalent to the conditions that $\det M\geq 0$ and $\Tr M>0$.   Thus, when these two conditions are satisfied, then the matrix $U$ can be chosen to be real and orthogonal, in which case the Takagi diagonalization of $M$ reduces to the standard diagonalization of a real symmetric matrix $M$ by a real orthogonal similarity transformation.

\end{appendices}

\end{document}

\clearpage